\documentstyle[11pt,epsfig,myext]{article}
\begin{document}

\def\uncatcodespecials{\def\do##1{\catcode`##1=12 }\dospecials}
\def\setupverbatim{\tt
  \def\par{\leavevmode\endgraf} \catcode`\`=\active
  \obeylines \uncatcodespecials \obeyspaces \parindent=5mm \parskip=0pt}
{\obeyspaces\global\let =\ } 
{\catcode`\`=\active \gdef`{\relax\lq}}
\def\beginverbatim{\par\begingroup\setupverbatim\doverbatim}
{\catcode`\|=0 \catcode`\\=12 
  |obeylines|gdef|doverbatim^^M#1\endverbatim{#1|endgroup}}

\include{definitions}

\begin{titlepage}

\begin{flushright}
{\rm Jagellonian University preprint TPJU-3/2004}\\
{\rm NSF-KITP-04-21}\\
{\rm HNINP-V-04-03}\\
\end{flushright}

\begin{center}
\vspace{1.0cm} 
  {\bf \LARGE Why do we need higher order fully exclusive} \\
\vspace{0.25cm} 
  {\bf \LARGE Monte Carlo generator for Higgs boson production} \\
\vspace{0.25cm} 
  {\bf \LARGE from heavy quark fusion at LHC? } \\
\end{center}
\vspace{2.0cm}

\begin{center}
  {\bf E. Richter-W\c{a}s$^{a,b}$, T. Szymocha$^b$ and Z. W\c{a}s$^{b}$ }\\ 
  \vspace{0.25cm}
  {\em $^{(a)}$ Institute of Physics, Jagellonian University, 30-059 Krakow, ul. Reymonta 4, Poland.}\\
  {\em $^{(b)}$ Institute of Nuclear Physics PAS, 31-342 Krakow, ul. Radzikowskiego 152, Poland.}\\
\end{center} 
 
\vspace{1.0cm}
\begin{center}
{\bf Abstract}
\end{center}
In  this paper we argue that having available higher order fully exclusive Monte Carlo 
generator for Higgs boson production from heavy quark fusion will be
mandatory for data analysis at LHC. The $H \to \tau \tau$ channel, a key
for early discovery of the Higgs boson in the MSSM scenario, 
is discussed.  
With simplified example and for $m_H$ = 120~GeV we show, that depending 
on choice among  presently 
available approaches, used for simulation  of Higgs boson production   
from $b \bar b H$ Yukawa coupling, final  acceptance for the signal events being reconstructed
inside mass window may differ by a factor of 3. The spread is even larger
(up to a factor of 10)  for other  production mechanisms (promising for 
some regions of the  MSSM parameter space).
The complete analysis, which necessarily will add stringent requirements for  
background rejection (such as identification of b-jet  or veto on  b-jet)
and which will require statistical combination of samples selected with 
different selection criteria may only enhance the uncertainty.


\vspace*{31mm}
\bigskip
\footnoterule
\noindent
{\footnotesize \noindent
Work supported in part by: the European Union 5-th Framework under contract HPRN-CT-2000-00149,
Polish State Committee for Scientific Research 
(KBN) grant 2 P03B 001 22, and US National Science Foundation under Grant No. PHY99-07949. 
}
\end{titlepage}

\section{Introduction}

The search for the Higgs boson is one of the primary task of the  
experiments at LHC \cite{:1999fr,unknown:1994pu}. Several studies discussed there,
concluded both for the Standard Model and Minimal Supersymmetric
Standard Model, that Higgs boson can be discovered at LHC, with the high 
margin of significance, 
for  the full  interesting range of its mass.

 One of the key signature for Higgs boson discovery, is with its decay into 
pair of $\tau$-leptons. 
For the MSSM Higgs scenario, decay into $\tau$-pair is strongly 
enhanced at  large $\tan \beta$, the parameter of the 
supersymmetric models. The promising 
(it depends on the Higgs boson mass and $\tan \beta$) production mechanisms 
are: the Higgs boson production
in association with the b-quarks, inclusive gluon-gluon fusion and,  for the lightest MSSM Higgs,
also vector-boson-fusion.

The analyses prepared for the MSSM Higgs boson searches in ATLAS,
presented in \cite{:1999fr},  rely mostly on the lepton-hadron 
decays of the $\tau$-pair, that is the case, when one of the $\tau$-leptons, 
from the Higgs boson decay, subsequently decays leptonically 
and another one hadronically
(hadron-hadron case was discussed only for the Higgs boson of large mass). 
The lepton-lepton channel contribution, to the Higgs boson 
discovery  potential, was  discussed in \cite{unknown:1994pu}, in context of the 
CMS detector.
 
On theoretical side,
the inclusive cross-section at large $\tan \beta$ is dominated by the bottom-quark fusion
process $b \bar b \to H$.  For the first time it was discussed in \cite{Dicus:1989cx}.
Recently, important progress has been achieved. The total cross-section 
for the  $b \bar b \to H$ process has been evaluated to the 
next-to-next-to-leading order (NNLO) \cite{Harlander:2003ai} 
in the so called  {\it variable flavour number scheme (VFS)} \cite{Dicus:1989cx,Dicus:1998hs,Maltoni:2003pn}. 
The result of this  NNLO calculation
shows almost no scale dependence. The inclusive  $b \bar b \to H$ cross-section
was obtained also at the next-to-leading (NLO), fixed order calculations for  the 
parton level process $gg, q \bar q \to b \bar b H$ \cite{Dittmaier:2003ej,Dawson:2003kb}.  There, the
{\it fixed flavour number scheme (FFS)} was used. 

Results obtained in these two schemes seem to be  compatible now with  each other, and show that
there is actually no large difference  between the NLO fixed order results
and use of the b-quark structure functions, if the proper factorisation scale
for this process is used \cite{Maltoni:2003pn}. Without the NNLO calculations \cite{Harlander:2003ai},
 understanding consistencies between FFS and VFS approaches would be on much weaker foundation.
This statement concludes what was 
discussed since a long time \cite{Dicus:1998hs, Maltoni:2003pn, Campbell:2002zm, Rainwater:2002hm}. 
The fixed order calculation shows a substantial scale dependence
and sufficient control of the residual large uncertainties, available only at higher 
orders, was necessary to compare results of the two approaches.

One should be well aware,
that for a given decay channel of $\tau$-pair, the statistical combination of 
the discovery evidence for events with one identified $b$-jet and events with no 
identified $b$-jet at all will be required, to achieve maximal (sufficient)  
sensitivity to the Higgs boson.
This defines, what are the relevant hard processes for the analyses, as designed in 
\cite{:1999fr}. For example 
if the identified final state includes single
 bottom quark (jet), then the relevant lowest order hard process in the VFS approach should be  
$gb \to bH$ \cite{Dicus:1998hs}. The cross-section for the $gb \to bH$ production has been 
also computed at NLO \cite{Campbell:2002zm} and the residual uncertainties due to the higher order
corrections are small. In the FFS scheme the relevant lowest order hard process will still be 
the $gg, q \bar q \to b \bar b H$.

At early stage of LHC operation, 
also statistical combination of signatures with different $\tau$ decay channels will need to be
performed. In addition, in the MSSM scenario,  
signals from different Higgs bosons (h, H and A) not degenerated in mass, 
might nonetheless overlap in the same mass window.  

Although the  NLO and even NNLO calculations became available for the 
integrated cross-sections, and impressive progress has been achieved in
understanding consistencies between (VFS)
and (FFS) approaches \cite{Harlander:2003ai,Dicus:1998hs,Maltoni:2003pn,Campbell:2002zm,Rainwater:2002hm},  
and even though some results on transverse Higgs momentum distribution are also discussed 
in the literature (see. e.g.~\cite{Boos:2003yi}),
only the {\it LO matrix element + parton shower}  approach is available
for the full event generation which allow for the subsequent detector simulation\footnote{
Here we refer explicitly to {\tt PYTHIA} or {\tt HERWIG} generators, which provide 
generation of $b \bar b \to H$, $gb \to bH$ and $gg \to b \bar b H$ in the lowest order
of the hard process only. We are not aware of any implementation
in form of the Monte Carlo event generator of the complete calculations
 of \cite{Harlander:2003ai}.}.

 
As we will attempt to illustrate in this paper, the foreseen by {\it experiments} signal reconstruction 
procedures are  very sensitive to the topology of the signal production process. Therefore, good 
understanding, from the 
theoretical perspective, of the Higgs production topologies will be mandatory.
This can be achieved only with fully exclusive higher order Monte Carlo generator, what we hope will be
visible from the present paper.

In this paper, as an example  we will use the SM-like 120 GeV mass
Higgs boson. We will compare reconstruction efficiencies
and final resolutions for the  different hard processes and for 
lepton-lepton and lepton-hadron $\tau$ decay modes.

Our paper is organized as follows. In section 2 we discuss Higgs boson production
using different hard processes and the appropriate cross sections. In sections 3 and 4 we review basic reconstruction and 
selection properties to be used by experiments. This in particular will explain how  idealized 
signatures translate into more realistic ones. In section 5 we collect numerical results 
for the $(\ell \ \ell \  p_T^{miss})$ signature, originating from the case when both $\tau$-leptons decay
leptonically. Similarly, in section 6 we collect numerical results 
for the $(\ell \ \tau$-jet $\ p_T^{miss})$ signature, originating from the case when one $\tau$ decays leptonically
and another one hadronically. Finally, conclusions, section 7 close the paper.

\section{Different hard processes}

Let us start discussion by presenting the naive Table~\ref{TS2.1} with cross-sections calculated for three
different hard processes: the (2 $\to$ 1) process $b \bar b \to H$, the (2 $\to$ 2) process $gb \to Hb$ and 
the (2 $\to$ 3) process $gg, q \bar q \to b \bar b H$, 
as obtained from {\tt PYTHIA 6.2} \cite{Sjostrand:2001yu} generator, 
according to its default initialisations\footnote{
These means choices for the QCD factorisation scale, minimum  bias model, parameters of the shower 
evolution, etc. etc. We think that, for example,   
discussion of consequences of the possible 
alternative choices is beyond the scope of this paper, it could only
dilute the aim of the paper. The appropriate theoretical framework, 
necessary for such a discussion need to be established first.
For the structure functions we have used CTEQ5L parametrisation.}.

Factor 4 difference in normalisation can be reported between $b \bar b \to H$ and
 $gg, q \bar q \to b \bar b H$ hard processes. 
The first one represents lowest order term 
in (VFS) scheme, second the lowest order term in the (FFS) scheme for the Higgs boson production
mediated by the $b \bar b H$ Yukawa coupling, which does not rely on the $b$-quark PDF's. 
Given recent clarification in \cite{Maltoni:2003pn} this difference could be minimised by using the proper 
factorisation scale. 

The question, whether differential distributions can impose significant effects on how
in experimental conditions, the Higgs boson signature is expected to be defined,
 was not adressed so far in theoretical calculations. The experimental studies
have shown clearly that the impact of the topological features of the production process can be significant 
on the overall efficiency of standard reconstruction procedure, see  ref.~\cite{:1999fr} pages: 746, 747.

One of the characteristic for event topology differential distribution is the 
transverse momenta of the Higgs boson, as it determines directly the average transverse
momenta and angular separation of the decay products.
In Fig.~\ref{FS2.1} we show Higgs boson transverse momenta distribution $p_T^{Higgs}$
as generated with different hard processes used.
The average $<p_T^{Higgs}>$ distribution for  $b \bar b \to H$  process
is  23~GeV,  for the $gb \to b H$ process is   31~GeV,  and finally 
for the $gg \to b \bar b H$ process is 28~GeV. 
The differences in event topologies seems, at a face value, not to be dramatic but we will study
nonetheless their impact further in the paper.

\begin{Tabhere} 
\newcommand{\lstrut}{{$\strut\atop\strut$}}
\begin{center}
\begin{tabular}{|c||c|} \hline \hline
Hard Process &  $\sigma \times BR$ [fb] \\
\hline \hline
$b \bar b \to H(\to \tau \tau)$  & 7.2  \\
\hline
$g b \to b H( \to \tau \tau)$ &  4.1    \\
\hline
$gg, q \bar q \to b \bar b H( \to \tau \tau) $ & 1.7    \\
\hline \hline
\end{tabular} 
\end{center} 
  \caption {\em Cross-section for signal production with b-quark
  Yukawa coupling. Results for there different hard processes  are collected.
  Branching ratios of $H \to \tau \tau$  is included. 
  The cross section is for 120 GeV mass SM Higgs boson at 14 TeV pp collision, 
  as generated with  default initialisations in {\tt PYTHIA}. 
\label{TS2.1}} 
\end{Tabhere}
\vspace{2.5mm}


\begin{Fighere}
\begin{center}
{
     \epsfig{file=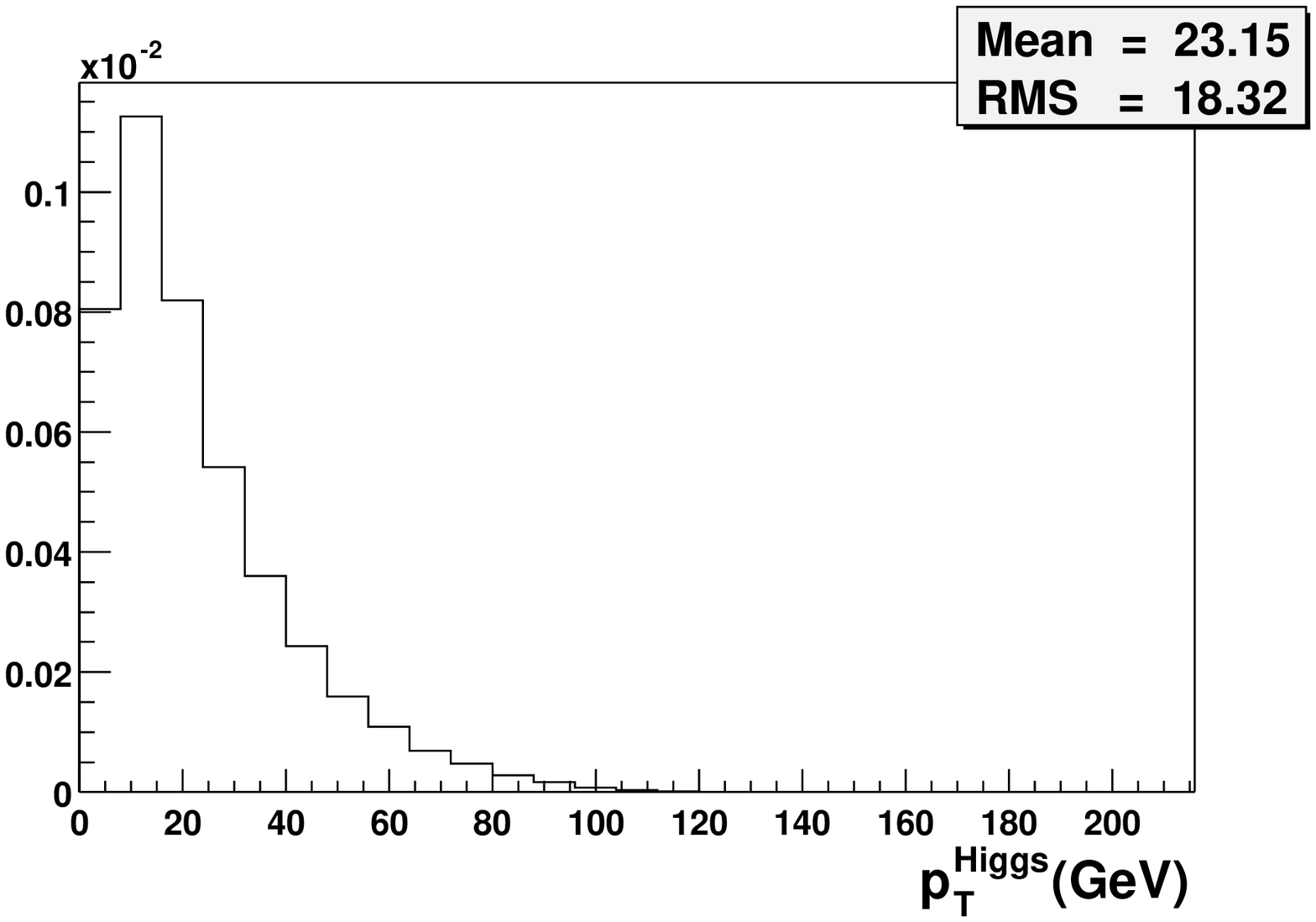,width=4.5cm, height=4.5cm}
     \epsfig{file=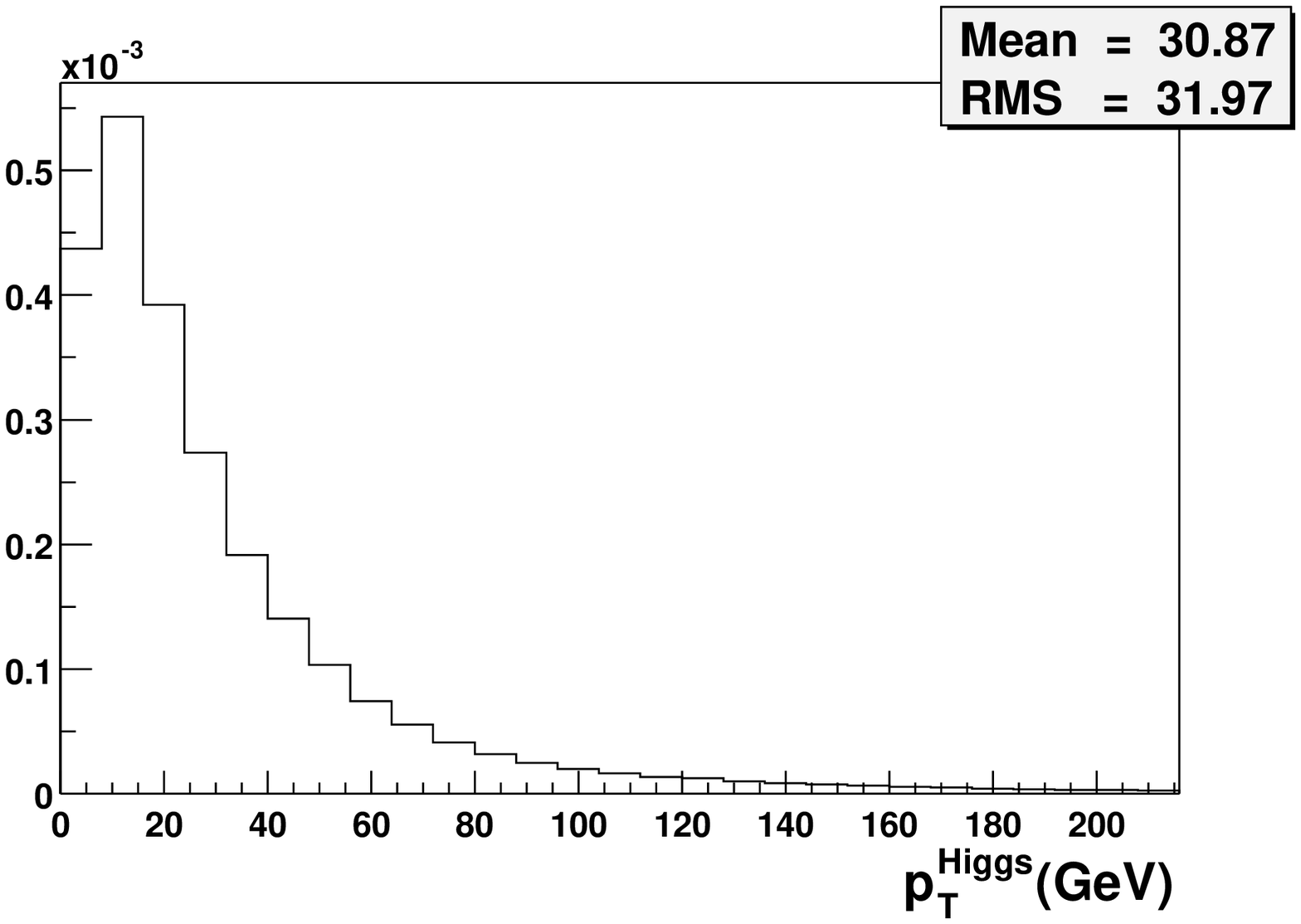,width=4.5cm, height=4.5cm}
     \epsfig{file=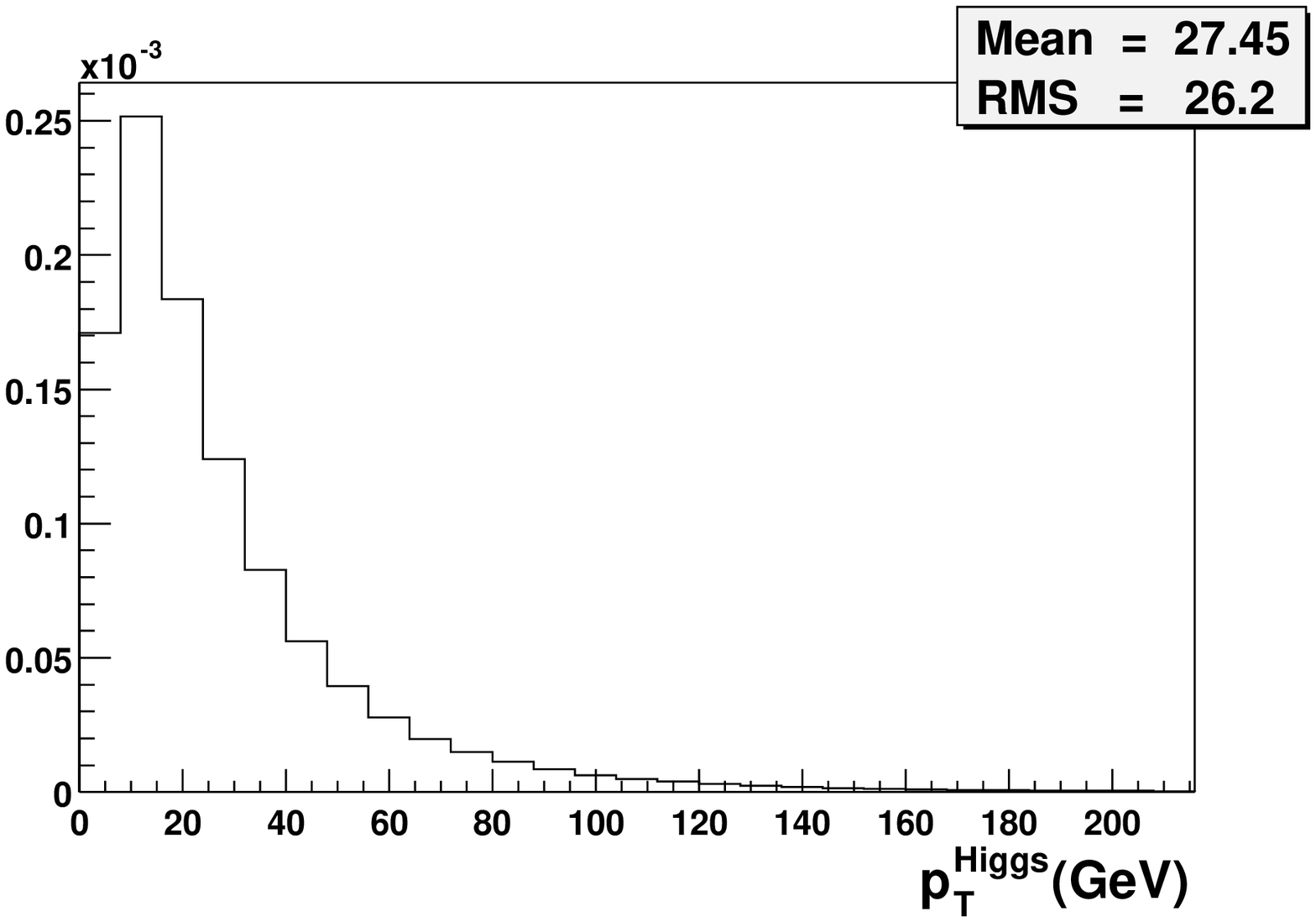,width=4.5cm, height=4.5cm}\\
}
\end{center}
\caption{\em
The $p_T^{Higgs}$ distribution for the three hard processes used:
$b \bar b \to H$ (left), $g  b \to b H$ (middle) and $gg \to b \bar b
H$  (right). Distribution normalised to total $\sigma \times BR$ (pb).
Each of these distribution is expected to include significant deficiencies. In the first case
generation of $p_T$ above certain value (max. shower scale) is not performed,
the later two, are valid only when respectively one/two final state $b$-quarks have 
$p_T$ above Sudakov turnover. These condition do not translate into constraint on  $p_T^{Higgs}$
in a straightforwad manner.
\label{FS2.1}} 
\end{Fighere} 
\vspace{2.5mm}

To perform realistic studies it was not enough to generate Higgs production process.
For that purpose, PYTHIA \cite{Sjostrand:2001yu} Monte Carlo was used. Essential
element consisted of simulating  the decays of the $\tau$-leptons. For that purpose 
 {\tt TAUOLA}~\cite{Jadach:1990mz,Jezabek:1991qp,Jadach:1993hs} 
and {\tt PHOTOS}~\cite{Barberio:1990ms,Barberio:1994qi} Monte Carlos were used. 
The three segments of  the generation were combined 
using interface presented in~\cite{Golonka:2003xt}. In this way,   
correct simulation of the  $\tau$-decays and the radiative photons emission 
from the Higgs boson decay cascade, relevant in particular for the hadron-lepton mode
were achieved. Also full spin correlation effects in the Higgs boson decays were assured.

\section{Reconstruction of the basic experimental signatures}

Reconstruction of the basic experimental signatures: leptons, $\tau$-jets, missing transverse energy,
have been done with the fast simulation of the simplified LHC detector, {\tt AcerDET}~\cite{Richter-Was:2002ch}. 
Although to large extend it is representing
the {\it best performance} of the detector, we believe that it is fairly 
adequate for the studies presented in this paper. The principle of  {\tt AcerDET} operation is
quite similar to the official fast simulation package of ATLAS Collaboration,
 {\tt ATLFAST} \cite{ATL-PHYS-98-131}. In contrary to the latter it relies on only those 
parameters of the detector which are available in the literature, eg \cite{:1999fr}, thus public. 
It is however missing regular internal updates of  {\tt ATLFAST}. 

The key experimental issues, which are relevant for discussed analyses can be 
shortly summarised as: $\bullet$
Both discussed channels have leptons ($e$ or $\mu$) in the final state and can be
triggered by either single- or the dilepton trigger. 
$\bullet$
The hadronically decaying tau's can be identified in the detector with good efficiency
and purity. The important ingredient in the $\tau$-jet identification
is the profile of the energy deposition in calorimeter and the number of tracks
pointing to the calorimeter cluster. 
$\bullet$
Missing transverse energy, $p_T^{miss}$,  is calculated after summing up all reconstructed visible signatures 
in the final states, including also cells and clusters in the calorimeter.

We have checked that if the {\tt ATLFAST} is used in simulation instead
of {\tt AcerDET}, a bit worse mass resolution and efficiency for the
$\tau$-jet reconstruction is obtained. This does not lead to changes
in  final conclusions of the paper.
For more detailed discussion on the comparison of full ATLAS detector simulation 
and {\tt AcerDET} results, relevant for the present analysis, 
see \cite{Richter-Was:2002ch}.

\section{Selection criteria}
 
Discussed here selection criteria are the minimal one for  triggering of a given event 
(at least one lepton in the final state) and for a good  reconstruction of the $\tau$-pair mass.
 Reconstruction of the invariant mass of the 
$\tau$-lepton system is  made under assumption
that the $\tau$-lepton is massless and decays collinearly\footnote{
We have checked that this assumption leads to about 2.5-3.5 GeV contribution to the gaussian width
of reconstructed invariant mass of the $\tau \tau$ system.}
 \cite{Aachen,:1999fr,unknown:1994pu}.
The procedure which have been used in \cite{:1999fr,unknown:1994pu} is equivalent, for 
the physical solutions, to the one used in ~\cite{Rainwater:1998kj,Plehn:1999xi} and described below.
Fractions of the two $\tau's$ momenta which are carried by measured visible
decay products, $x_{\tau_1}$, $x_{\tau_2}$,  can be calculated from solving equations of
conservation of the transverse momenta components of the Higgs boson, 
reconstructed from the visible products (leptons, $\tau$-jets) and missing 
transverse energy. The physical solutions
(events with resolved neutrinos) are those for which $0 < x_{\tau_1}, x_{\tau_2} < 1$.  
For events with physical solution, the invariant mass of the system of the
visible decay products of $\tau$'s is calculated and the invariant mass 
of the $\tau$-system is expressed as  $m_{\tau \tau} = m_{vis} / \sqrt{x_{\tau_1} \cdot x_{\tau_2}}$.

We start with selection procedure as defined in \cite{:1999fr} and refined in \cite{SN-2003-024}
for the lepton-lepton final state. The {\it basic selection}
consists of requiring: two isolated leptons  within acceptance of the detector;
threshold on the minimal angular separation between leptons; threshold on the
reconstructed missing transverse energy and  kinematics of events which allow
to resolve equations for neutrinos momenta, i.e. for invariant mass of the $\tau$-pair system
to be calculable.
The {\it additional  selection}, also introduced already in \cite{:1999fr, SN-2003-024}, is necessary 
to optimise  resolution of the reconstructed
invariant mass of the $\tau$-lepton pair.

Slightly modified selection is used for lepton-hadron final state, the {\it basic selection}  
criteria have to be  adjusted, already from the beginning, 
to reject expected  reducible backgrounds ($t \bar t$, $W+j$).

For the explicit definition of {\it basic selection} and {\it additional selection}
 cuts\footnote{
Angle $\phi$ is measured in the plane transverse to the beam axis. The $R_{\ell \ell}, R_{\ell \tau-jet} $ 
denote cone separation, expressed by the difference in pseudorapidity $\eta$ and angle $\phi$,
 $R = \sqrt{ (\Delta \phi)^2 + (\Delta \eta)^2}$. The transverse mass in lepton-hadron channel is calculated as
$m_{\ell, miss} = \sqrt{ (p_T^{\ell})^2 + (p_T^{miss})^2}$.
For more comments on definition of variables see  \cite{SN-2003-024}.}
see later: Tables \ref{TS5.1} and \ref{TS6.1} respectively  in sections 5 and 6.

\section{The $(\ell$~$\ell$~$p_T^{miss})$ final state}

In Fig.~\ref{FS5.1} we show distributions of the kinematical variables in the transverse plane:
$\sin(\Delta \phi_{\ell \ell})$, $\; p_T^{miss}$, $\; \cos(\Delta \phi_{\ell \ell})$,
used for event selection.
One can observe some differences in the shape of the 
distributions, which do not seem very significant (except  for $\cos(\Delta \phi_{\ell \ell})$), 
but nevertheless  
lead to the cumulated effect on the acceptances of order of a factor 3 (see Table~\ref{TS5.1})
for signal events in lepton-lepton mode, generated with 
different LO 
processes. The very initial acceptances are comparable (first line of  Table~\ref{TS5.1}),
divergences start with adding selection and reconstruction requirements.

\begin{Tabhere} 
\newcommand{\lstrut}{{$\strut\atop\strut$}}  
\begin{center}
\begin{tabular}{|c||c|c|c||c|c| } \hline \hline
Selection & $ b \bar b \to H$ &  $ g b \to b H$ &   $ b \bar b H$ & $ gg \to H$ & $qqH$   \\
\hline \hline 
Basic selection &  & & & &   \\
\hline \hline 
 2 iso $\ell$, $p_T^{\ell}>15$ GeV & $18.6 \cdot 10^{-2}$ & $18.4\cdot 10^{-2}$ & $ 19.2 \cdot 10^{-2}$ 
 &  $ 18.6 \cdot 10^{-2}$& $ 21.3 \cdot 10^{-2}$ \\
\hline \hline
$|sin(\Delta \phi_{\ell \ell})| > 0.2$ & $ 9.3 \cdot 10^{-2}$  & $10.1 
\cdot 10^{-2}$  & $ 10.1 \cdot 10^{-2}$  &  $ 10.4 \cdot 10^{-2}$&$ 19.0 \cdot 10^{-2}$  \\
\hline
$p_T^{miss}> 15$ GeV & $ 5.5 \cdot 10^{-2}$   & $7.0 \cdot 10^{-2}$ & $ 6.4 \cdot 10^{-2}$   &$ 7.9 \cdot 10^{-2}$  & $ 17.7 \cdot 10^{-2}$ \\
\hline
resolved neutrinos &  $ 4.4 \cdot 10^{-2}$  & $5.8 \cdot 10^{-2}$ & $ 5.1 \cdot 10^{-2}$   & $ 7.0 \cdot 10^{-2}$& $ 16.5 \cdot 10^{-2}$\\
\hline \hline 
Additional selection &  & &  & &  \\ 
\hline \hline 
$p_T^{miss}> 30$ GeV &  $ 1.4 \cdot 10^{-2}$  &  $3.0 \cdot 10^{-2}$ & $ 2.2 \cdot 10^{-2}$  & $ 4.2 \cdot 10^{-2}$  &  $ 13.4 \cdot 10^{-2}$  \\
\hline
$cos(\Delta \phi_{\ell \ell}) > -0.9$ &  $ 1.0 \cdot 10^{-2}$  &  $ 2.5
\cdot 10^{-2}$ & $ 1.8 \cdot 10^{-2}$   &$ 3.8 \cdot 10^{-2}$ &$ 12.8 \cdot 10^{-2}$ \\  
\hline
$R_{\ell \ell} < 2.8$ &  $ 9.5 \cdot 10^{-3}$  & $2.5 \cdot 10^{-2}$  & $ 1.8 \cdot 10^{-2}$  &$ 3.7 \cdot 10^{-2}$  & $ 12.7 \cdot 10^{-2}$ \\  
\hline \hline
\end{tabular}
\end{center}
  \caption {\em The cumulative acceptances of the selection criteria for different
approaches of modeling production process. For each subsequent line effect of the
additional cut off is added. 
\label{TS5.1}} 
\end{Tabhere}
\vspace{2.5mm}

From the  Table~\ref{TS5.1} we can read, that a large systematic theoretical uncertainty
must be assumed for the efficiency of the selection and reconstruction procedure,
as a consequence of the different choices in modeling of the production process topology. 
For illustration, we include results for the other production processes\footnote{
For the production processes $gg \to H$ and $qqH$, which we will also include 
later in our discussion of acceptance, default initialisations of {\tt PYTHIA 6.2} are used as well.
}: gluon-fusion  $gg \to H$, 
and vector-boson-fusion, $qq \to qqH$ as well.
These processes will be included
in the complete analysis of the Higgs signal observability as the channels are expected 
to be promising for some regions of the MSSM parameter space \cite{Rainwater:1998kj,Plehn:1999xi}.
 The final spread on the efficiency is then even larger.

\begin{Fighere}
\begin{center}
{ 
     \epsfig{file=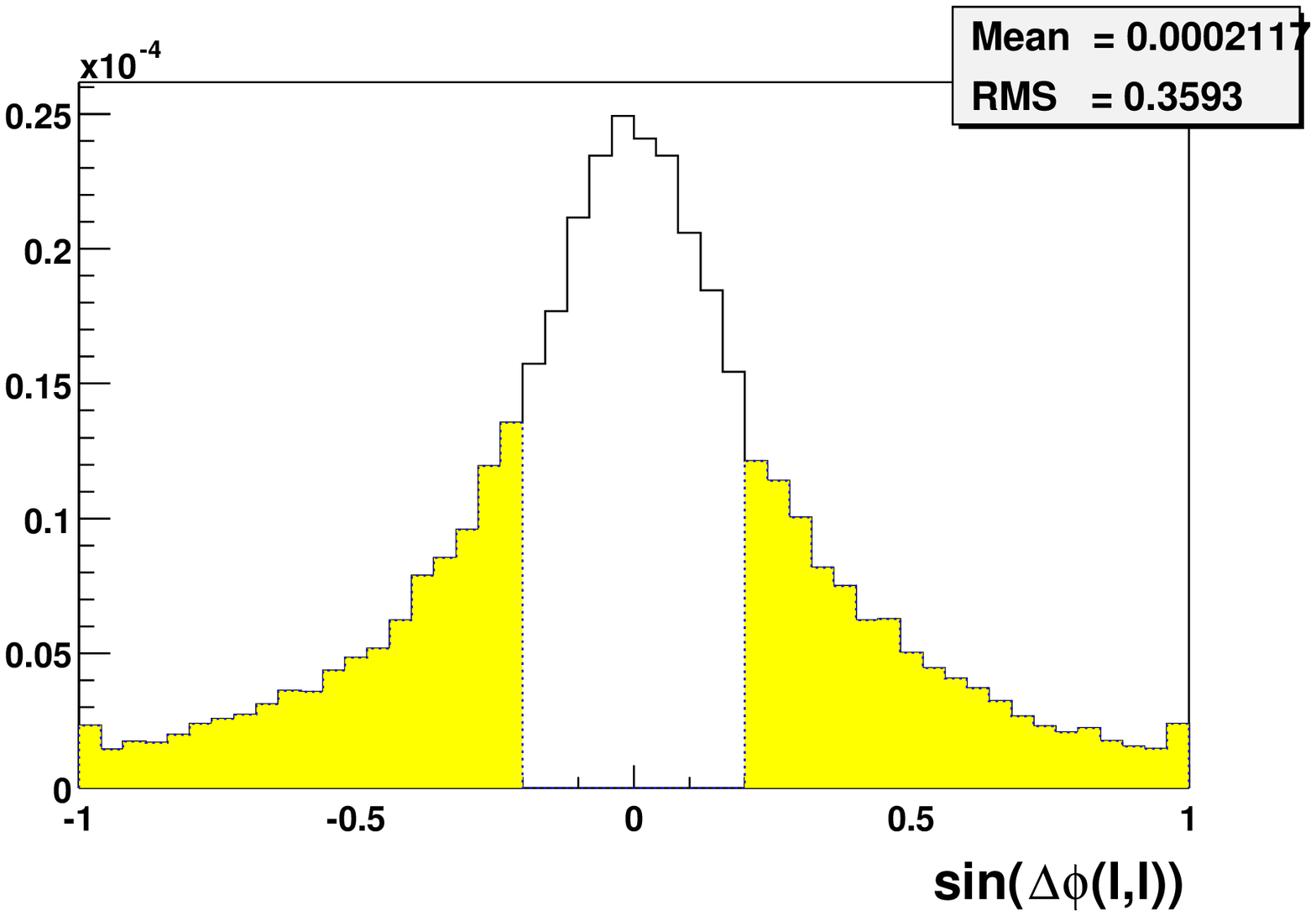,width=4.5cm, height=4.5cm}
     \epsfig{file=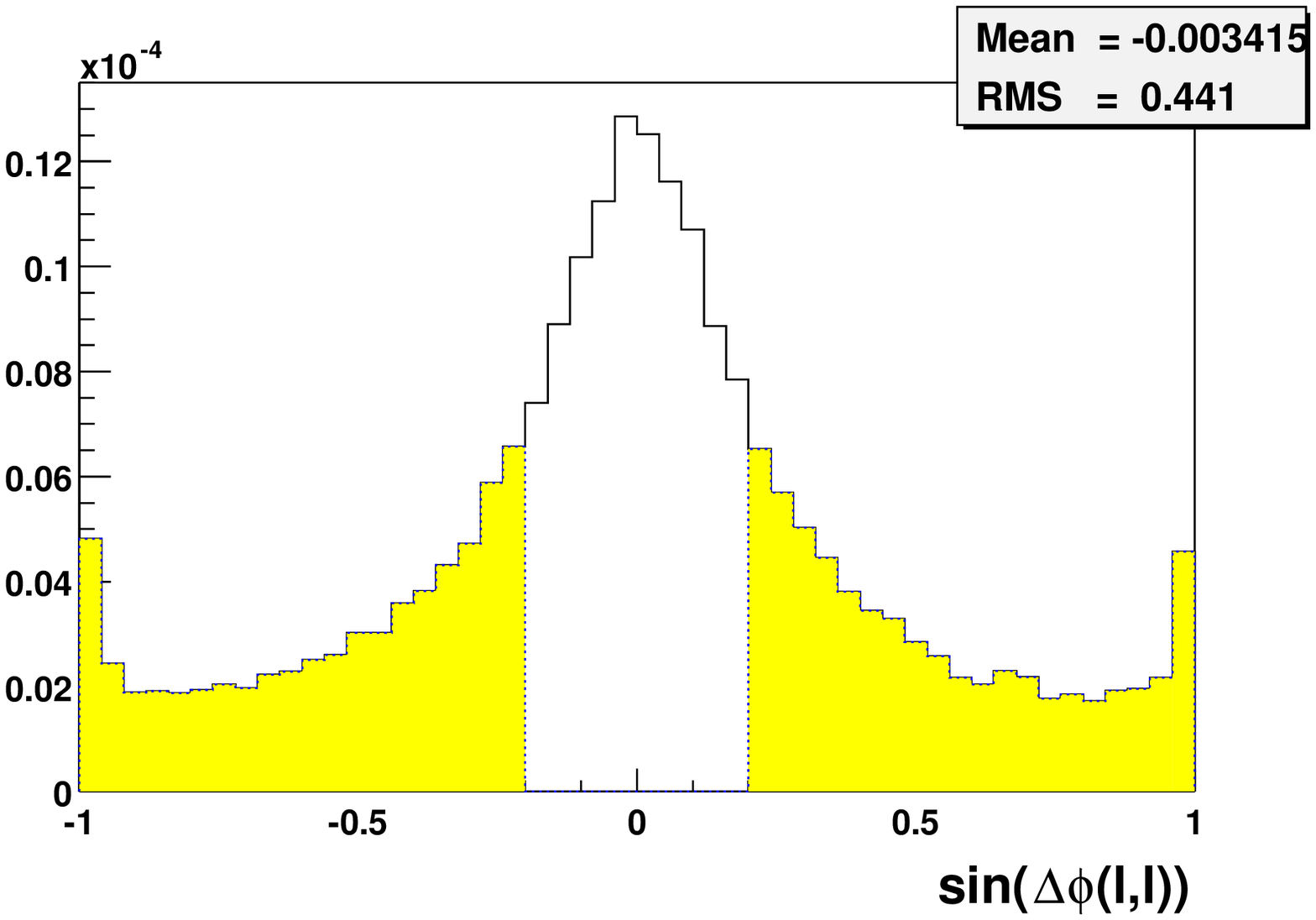,width=4.5cm, height=4.5cm}
     \epsfig{file=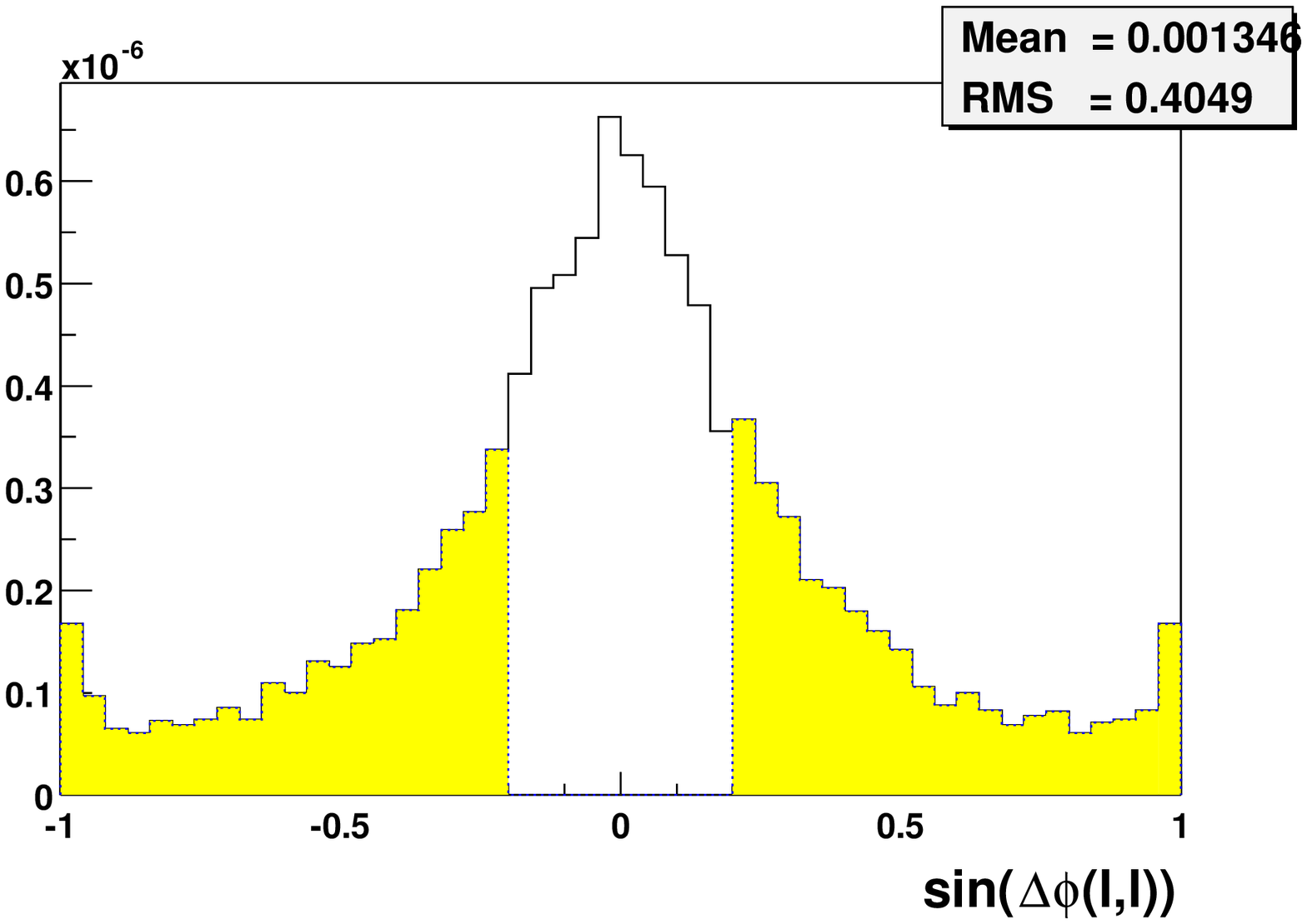,width=4.5cm, height=4.5cm}\\
}
{
     \epsfig{file=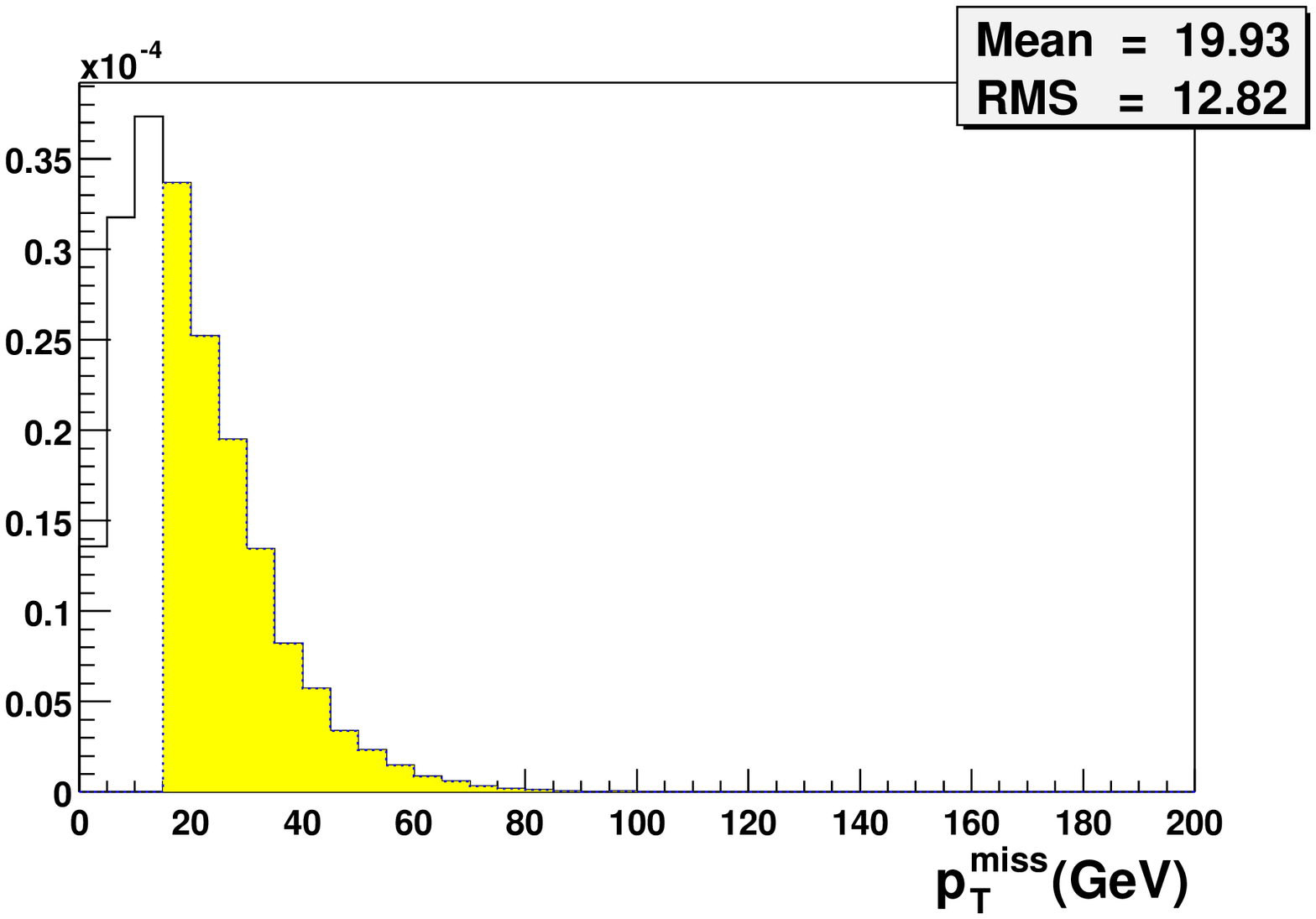,width=4.5cm, height=4.5cm}
     \epsfig{file=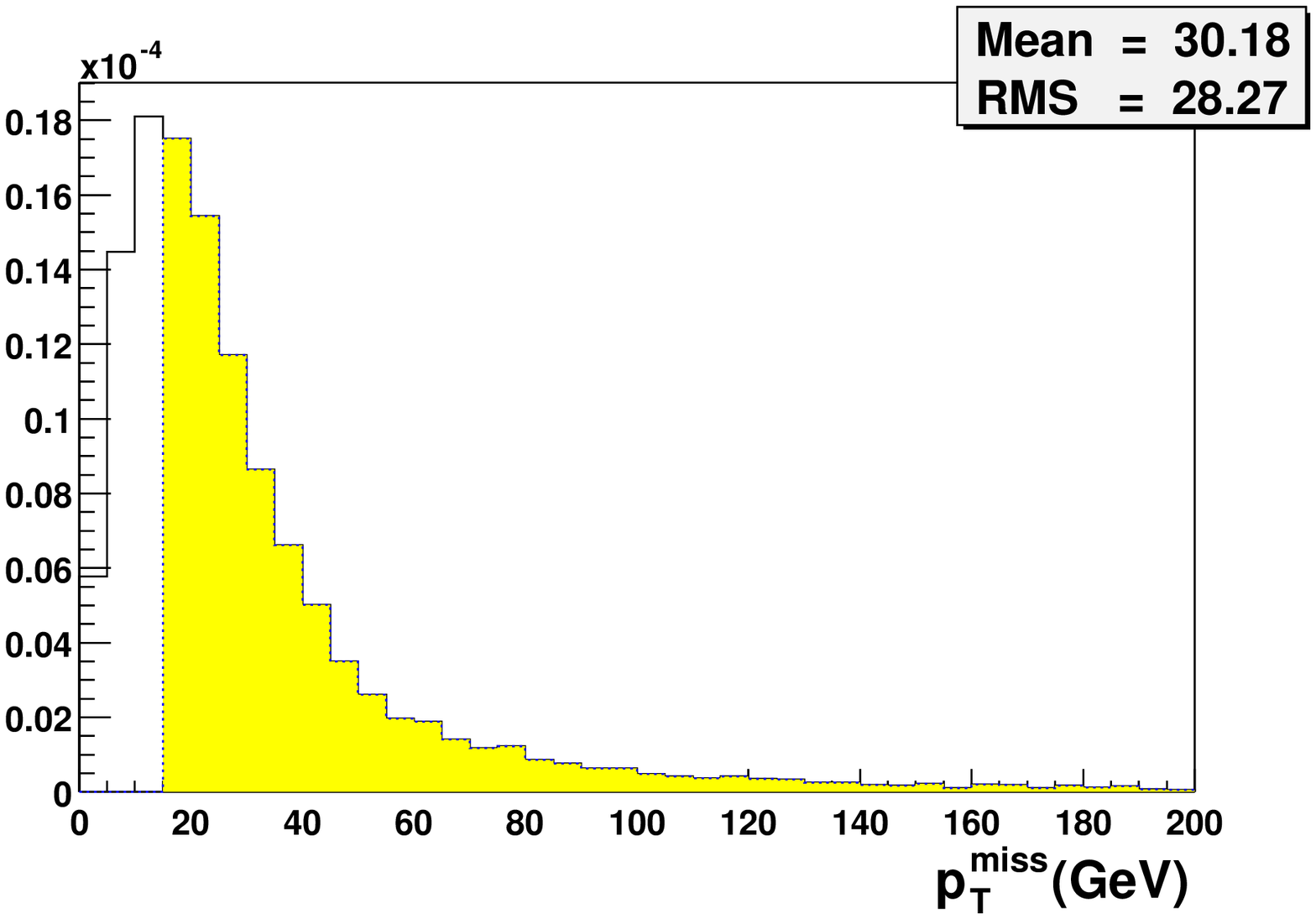,width=4.5cm, height=4.5cm}
     \epsfig{file=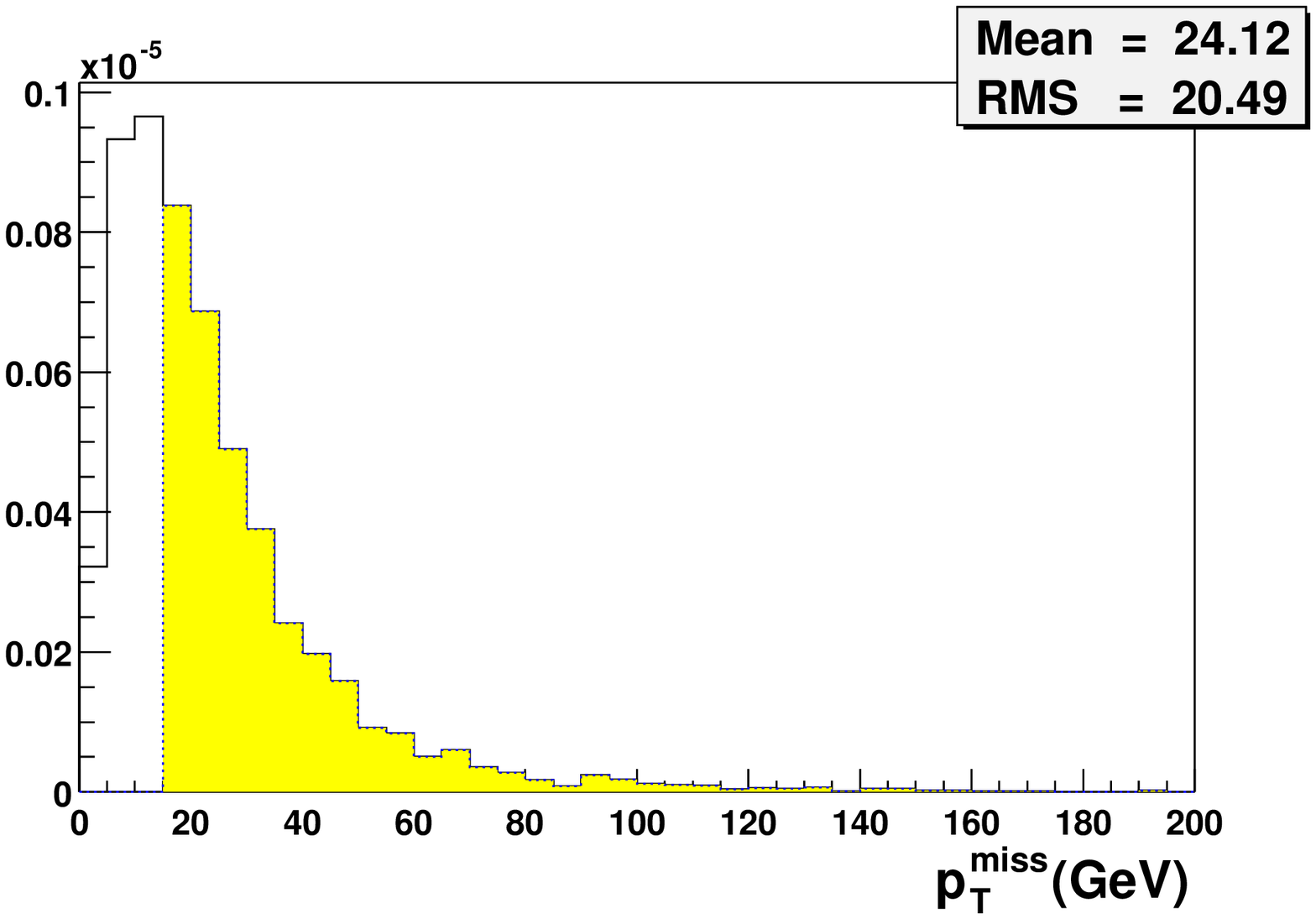,width=4.5cm, height=4.5cm}\\
}
{
     \epsfig{file=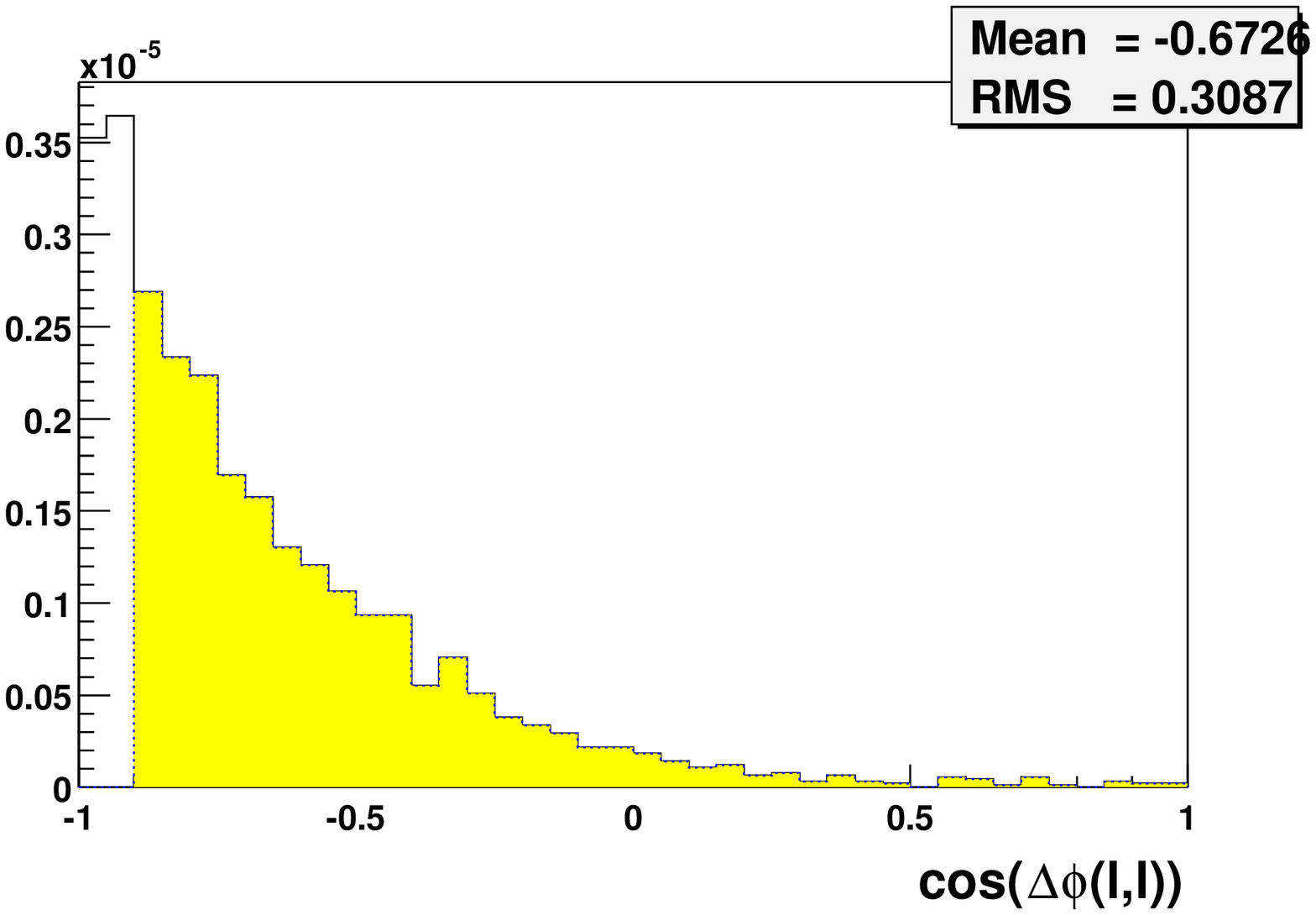,width=4.5cm, height=4.5cm}
     \epsfig{file=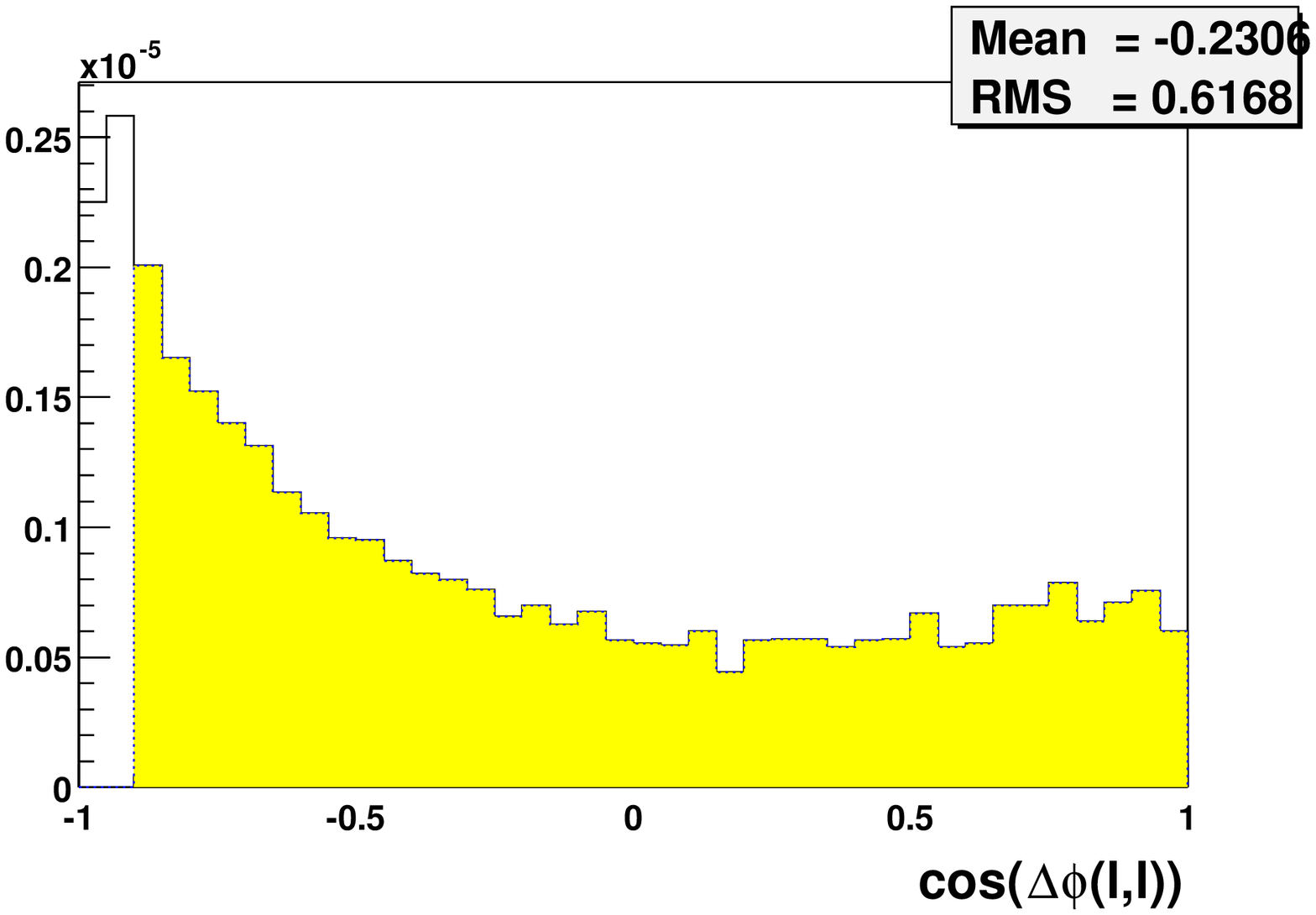,width=4.5cm, height=4.5cm}
     \epsfig{file=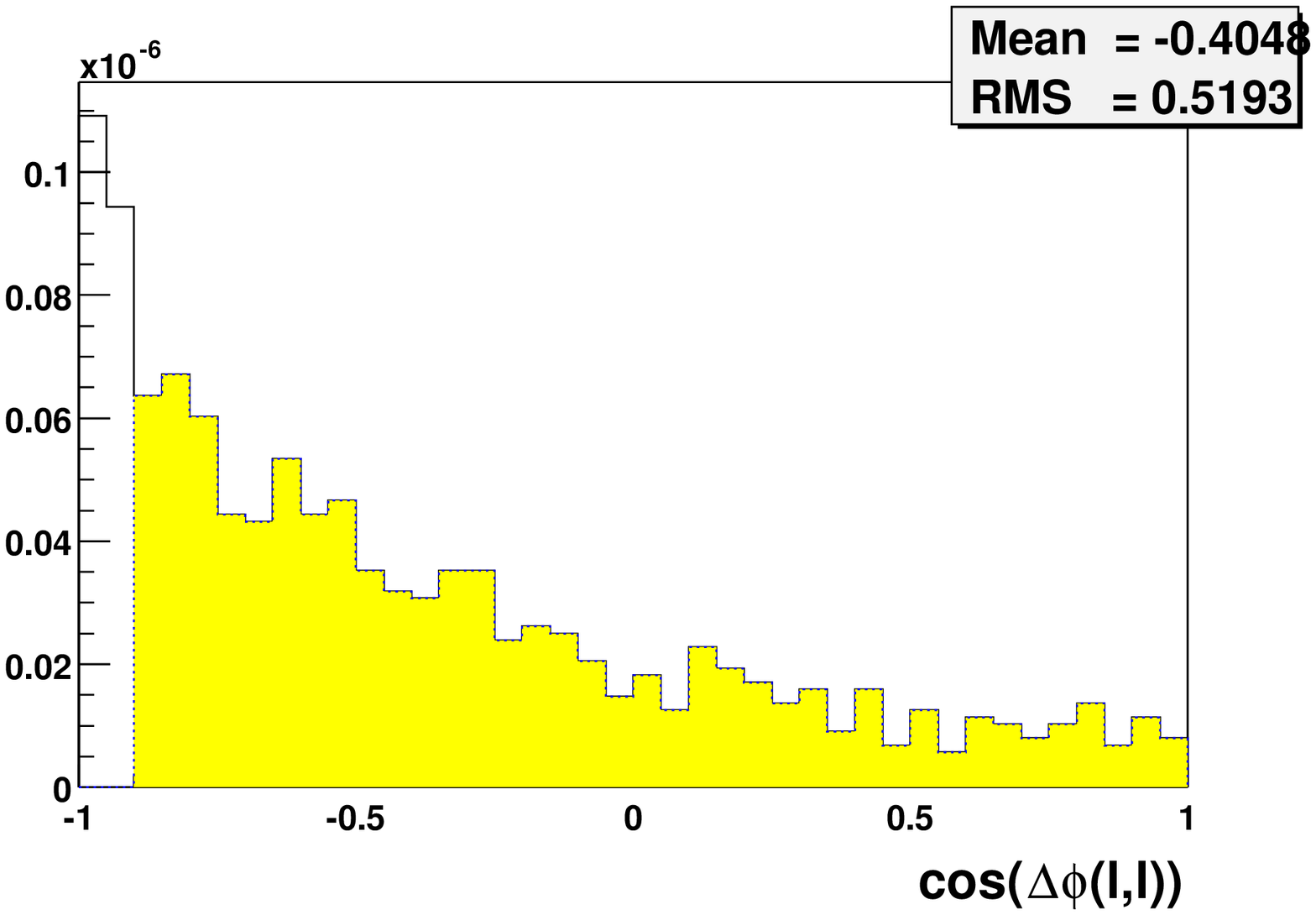,width=4.5cm, height=4.5cm}\\
}
\end{center}
\caption{\em
The characteristic kinematical distributions before respective selection, for 
different hard processes used: $b \bar b \to H$ (left column), $gb
\to b  H$ (middle column) and $q \bar q \to b \bar b H$ (right
column). The shaded (yellow) area will be accepted by selection.
Distributions normalised to total $\sigma \times BR$ [pb]. Every line of the plots
correspond to variables as used in cuts: lines 2, 3 and 6 of table \ref{TS5.1}.
Symmetric shape of the distributions of the first line of the plots provides
technical test of the simulation.    
\label{FS5.1}} 
\end{Fighere}

\vspace{2.5mm}

\begin{Tabhere} 
\newcommand{\lstrut}{{$\strut\atop\strut$}}  
\begin{center}
\begin{tabular}{|c||c|c|c||c|c|} \hline \hline
Selection & $ b \bar b \to H$ &  $g b \to b H$ & $ b \bar b H$  & $ gg \to H$ & $qqH$  \\
\hline \hline 
Basic selection & 17.3 GeV & 14.1 GeV & 16.0 GeV & 11.6 GeV & 8.8 GeV  \\
                & (63.6 \%)  &(68.6 \%) & (67.4\%) & (79.1\%)&  (88.7\%)\\
\hline
$p_T^{miss}> 30$ GeV &  15.8 GeV & 11.9 GeV    & 13.0 GeV & 9.8 GeV & 8.3 GeV  \\
                &  (73.0 \%)  &( 77.7 \%) & (76.9\%)& (86.5\%) &(91.3\%) \\
\hline
$cos(\Delta \phi_{\ell \ell}) > -0.9$ & 14.5 GeV & 11.4 GeV  & 12.3 GeV  & 9.3 GeV  & 8.1 GeV   \\  
                &  (80.8 \%)  &( 82.6 \%)  & (82.0\%) & (90.0\%)  & (93.1\%) \\
\hline
$R_{\ell \ell} < 2.8$ & 14.4 GeV & 11.4 GeV & 12.1 GeV & 9.2 GeV & 8.1 GeV   \\  
                &  (81.6 \%)  &( 83.2 \%) &(82.9\%)  & (90.6\%) & (93.4\%) \\
\hline \hline
\end{tabular}
\end{center}
  \caption {\em Resolution of the Gaussian fit to the reconstructed invariant mass of the $\tau \tau$
system for different approaches of modeling production process.
In brackets is shown acceptance within
mass window of $m_H~\pm$~20~GeV. 
\label{TS5.2}} 
\end{Tabhere}
\vspace{2.5mm}

\begin{Fighere}
\begin{center}
{
     \epsfig{file=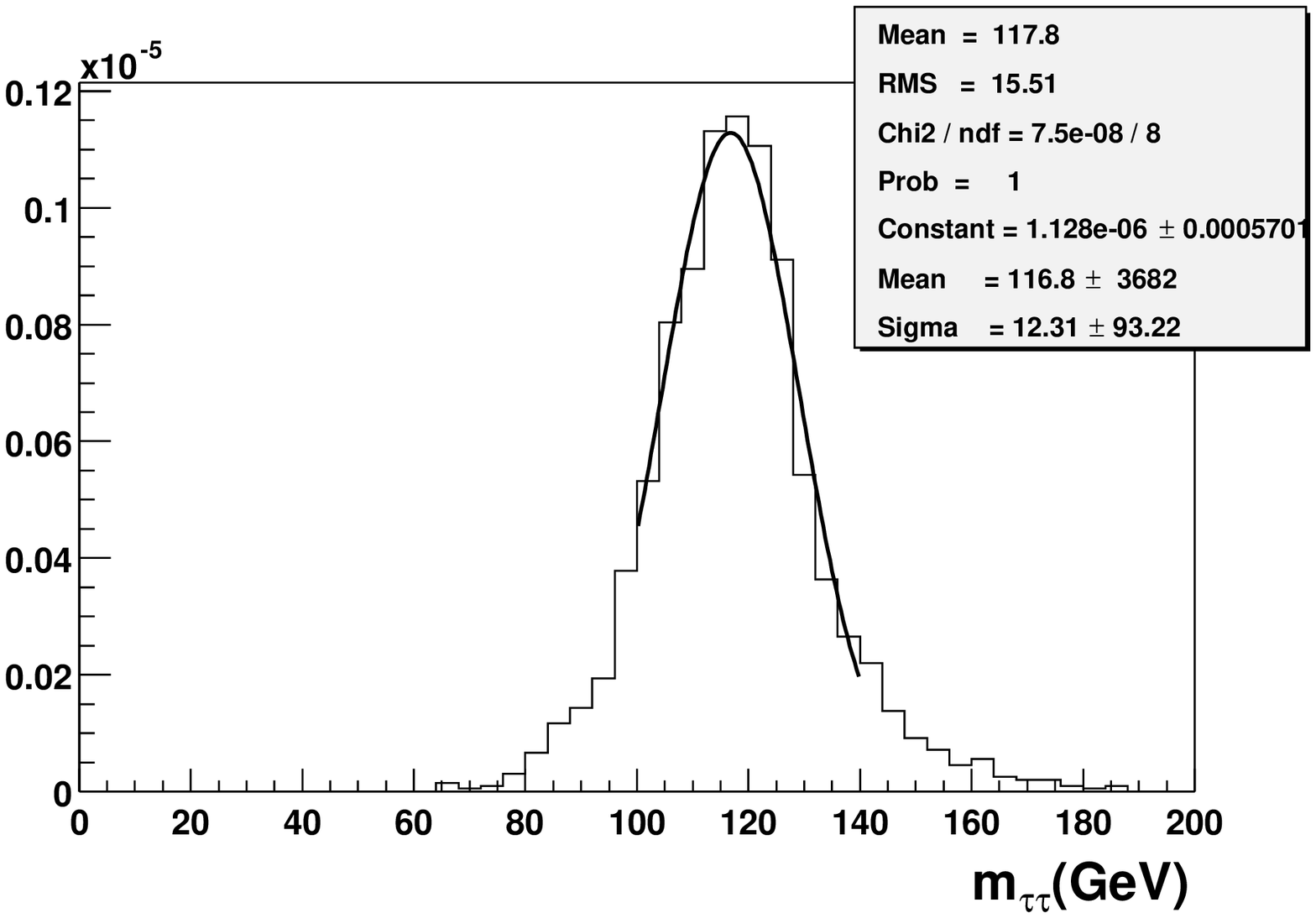,width=6.5cm, height=6.0cm}
}
\end{center}
\caption{\em
The reconstructed mass of the $\tau's$ pair, for the 
$gg \to b \bar b  H$  events.
Distribution after $cos(\Delta \phi_{\ell \ell})$ selection cut off is shown. Only those
events which are reconstructed in constrained  mass window, eg. $m_H~\pm$~20~GeV,
will be considered as contributing to the Higgs signature.
\label{FS5.2}} 
\end{Fighere}

In Table~\ref{TS5.2} we compare  resolution of the Gaussian fit to the reconstructed 
invariant mass of the $\tau$-pair system, and acceptances inside mass window of
$m_H~\pm$~20~GeV. The natural width is negligible for the Higgs boson with 120~GeV 
mass, the  non-zero resolution spread comes exclusively from the reconstruction procedure 
(assumptions of the collinearity in the $\tau$'s decay, and $p_T^{miss}$ reconstruction).
As an example, in Fig.~\ref{FS5.2} we show reconstructed invariant mass 
for the $gg \to b \bar b H$ events.
Resolutions for $b \bar b \to H$ and $ gb \to bH$ topologies (Table~\ref{TS5.2}) differ up to 30\%.
 This additional effect  will enhance the impact of choice
of the hard process  on the total
acceptance, on top of the one, introduced by  the differences in the selection
 efficiencies discussed in Table~\ref{TS5.1}.

Estimates for the total cumulative acceptance inside mass window,
as presented in  Table~\ref{TS5.3}, give clear indication of the size
of the systematic uncertainty which should be (at present) assigned to the predictions
on the expected number of signal events.

Note that the very initial acceptances 
(first line(s) in Table~\ref{TS5.1}) were comparable for all discussed processes. 
Then the selection was designed for signal reconstruction only. 
Additional
features of topologies, like presence of extra hard partons was not explored yet.
The difference for the total cumulated acceptances, which arised after additional selection
necessary for the signal reconstruction 
{\it inside}  mass window, is a factor of 3 for $b \bar b \to H$ versus $gb \to bH$ 
topologies. It is even a factor of 4 for $gg \to H$ versus $b \bar b \to H$ topologies.
The  {\it additional selection } as it was already pointed out,  is necessary for  
resolution of the reconstructed invariant mass of the $\tau$-lepton pair.

\begin{Tabhere} 
\newcommand{\lstrut}{{$\strut\atop\strut$}}
\begin{center}
\begin{tabular}{|c||c|c|c||c|c|} \hline \hline
Selection & $ b \bar b \to H$ &  $g b \to b H$ & $ b \bar b H$ & $ gg \to H$ & $qqH$ \\
\hline \hline 
Basic selection &  $ 2.8 \cdot 10^{-2}$ &  $ 4.0 \cdot 10^{-2}$ &  $ 3.4 \cdot 10^{-2}$  &  $ 5.6 \cdot 10^{-2}$  &  $ 14.6 \cdot 10^{-2}$ \\
\hline
$p_T^{miss}> 30$ GeV &  $ 1.0 \cdot 10^{-2}$ &  $  2.3 \cdot 10^{-2}$  &  $ 1.7 \cdot 10^{-2}$ &  $ 3.7 \cdot 10^{-2}$ &  $ 12.2 \cdot 10^{-2}$    \\
\hline
$cos(\Delta \phi_{\ell \ell}) > -0.9$ &  $ 8.2 \cdot 10^{-3}$  &  $
 2.1 \cdot 10^{-2}$  &  $  1.5 \cdot 10^{-2}$   &  $ 3.4 \cdot 10^{-2}$ &  $ 12.0 \cdot 10^{-2}$ \\  
\hline
$R_{\ell \ell} < 2.8$ &  $ 7.8 \cdot 10^{-3}$ &   $ 2.1 \cdot 10^{-2}$  &   $ 1.5 \cdot 10^{-2}$& 
 $ 3.4 \cdot 10^{-2}$ &   $ 11.9 \cdot 10^{-2}$  \\  
\hline \hline 
\end{tabular}
\end{center}
  \caption {\em The cumulative acceptance in the mass window $m_H~\pm$~20~GeV,
for different approaches of modeling production process.
\label{TS5.3}} 
\end{Tabhere}
\vspace{2.5mm}

\section{The $\ell$~$\tau$-jet ~$p_T^{miss}$ final state}

Let us now turn to the second case, where one of the $\tau$'s decays hadronically. We will proceed
similarly as in the previous section. We start with
Fig.~\ref{FS7.1} where  we show distributions of the kinematical variables used later
for  events selection:
$\sin(\Delta \phi_{\ell \ \tau-jet})$, $m_T^{\ell \ miss}$, $\cos(\Delta \phi_{\ell \ \tau-jet})$. 
One can observe some differences in the shape of these distributions, depending
on the choice of the  hard process (similarly as for lepton-lepton channel,
the $\cos(\Delta \phi_{\ell \ \tau-jet})$ is the
most sensitive to this choice). The cumulated effect on the acceptances
is of the order of a factor of 3 in this case (see Table~\ref{TS6.1}) similarly 
as it was in the lepton-lepton channel. 

The Gaussian resolution for the reconstructed invariant mass of the $\tau \tau$ system mass is specified in 
Table~\ref{TS6.2}. Obtained resolution on average is 10\% better than for the lepton-lepton channel,
which might seem contrary to the intuition (we need to reconstruct $\tau$-jet, what is less precise than
reconstruction of a lepton). 
The contradiction  is due to the fact that the $\tau$-jet spectrum 
from $\tau \to had \ \nu_{\tau}$ decay is harder than the lepton spectrum from
 $\tau \to \ell \nu_{\tau} \nu_{\ell}$
decay (in the second case we have two neutrinos).
Thus, in the first case the assumption of the collinear decay for $\tau$ lepton works better for the invariant 
mass reconstruction, simply because there is less of neutrino energy.

Estimates for the total cumulative acceptance inside mass window,
presented in  Table~\ref{TS6.3}, give clear indication of the size
of the systematic uncertainty which should be assigned to the predictions
for the expected number of signal events. 
The difference for the total cumulated acceptance for the signal reconstruction 
inside mass window are similar in size to the one in the lepton-lepton final state.

\begin{Fighere}
\begin{center}
{ 
     \epsfig{file=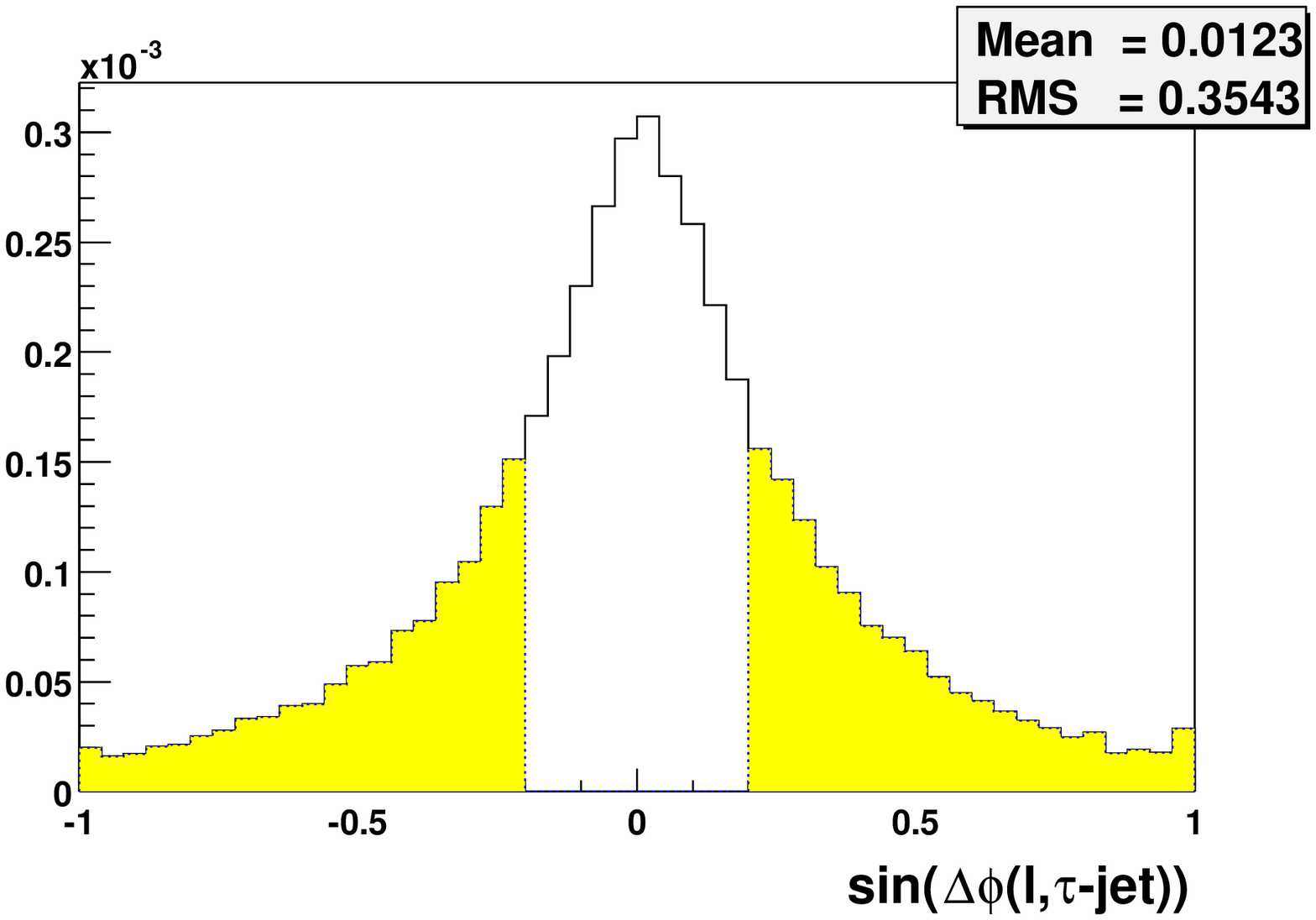,width=4.5cm, height=4.5cm}
     \epsfig{file=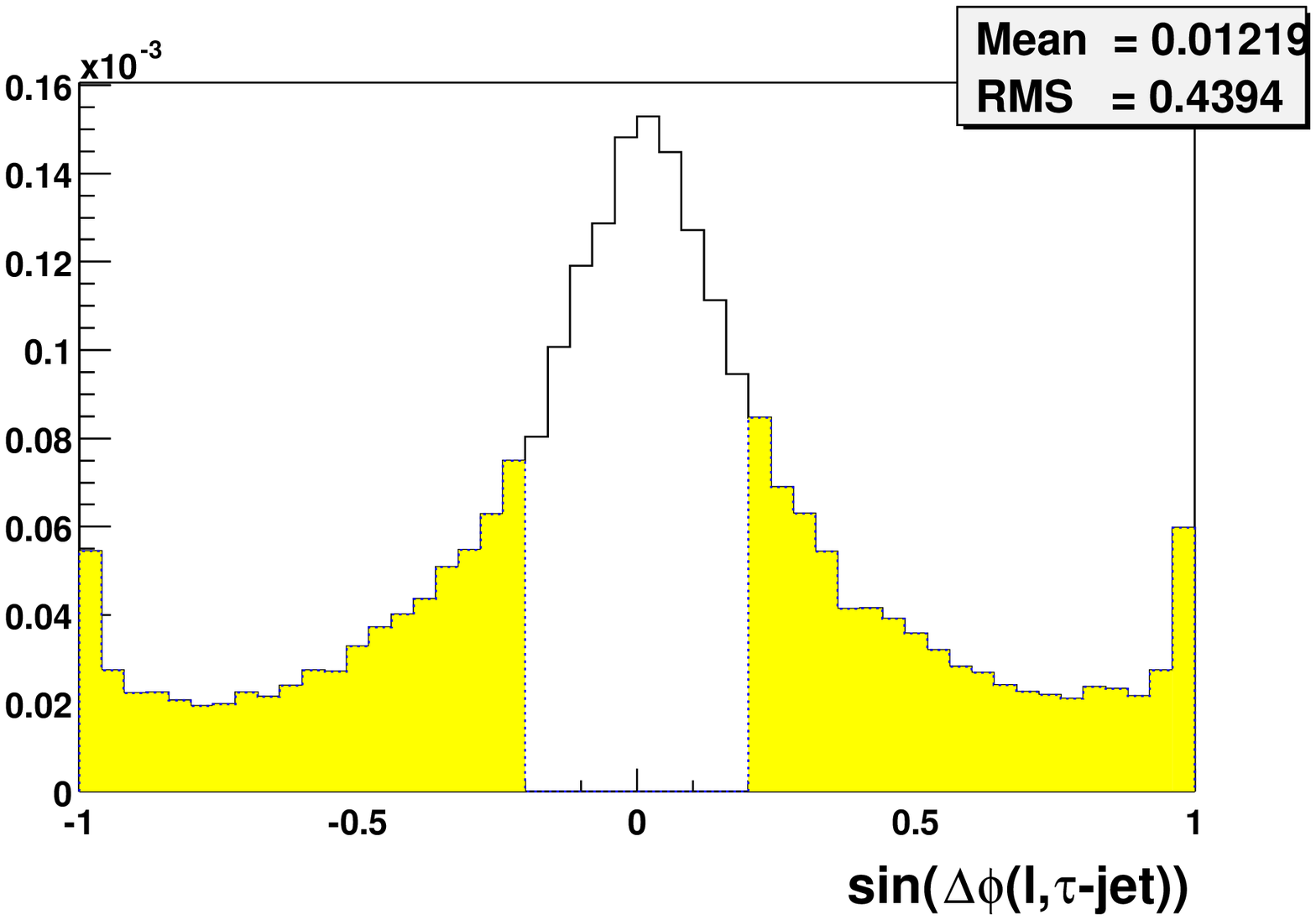,width=4.5cm, height=4.5cm}
     \epsfig{file=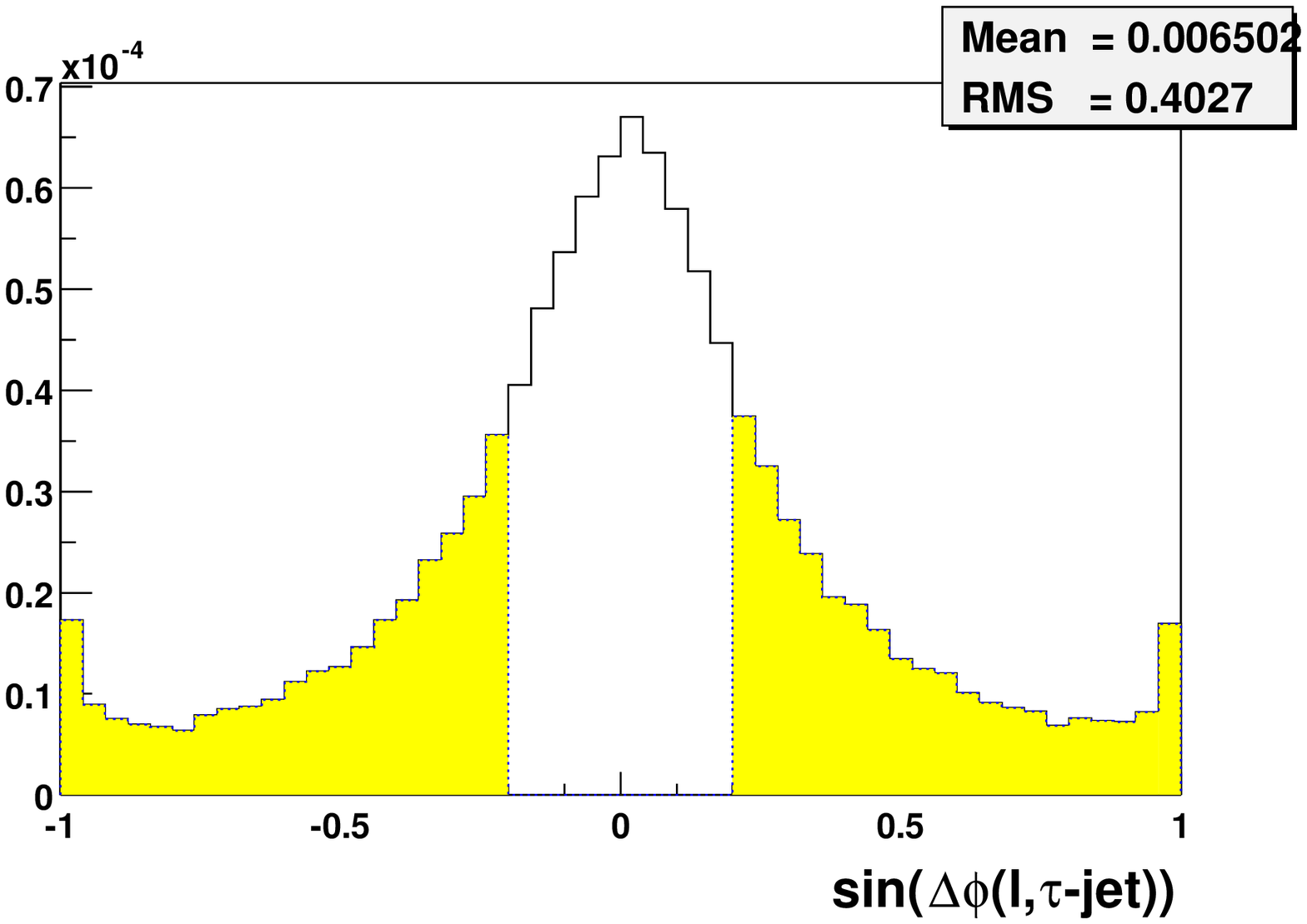,width=4.5cm, height=4.5cm}\\
}
{
     \epsfig{file=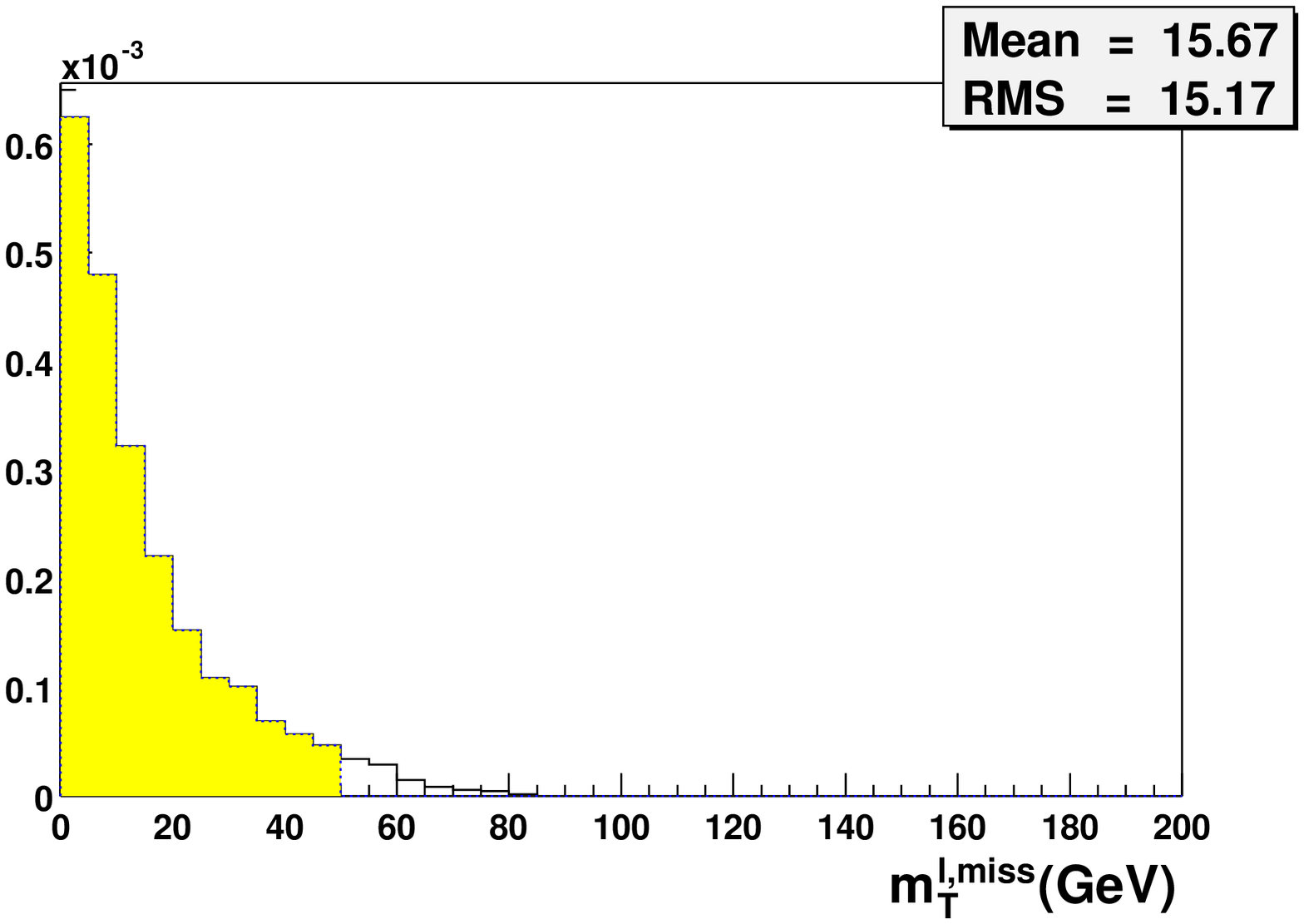,width=4.5cm, height=4.5cm}
     \epsfig{file=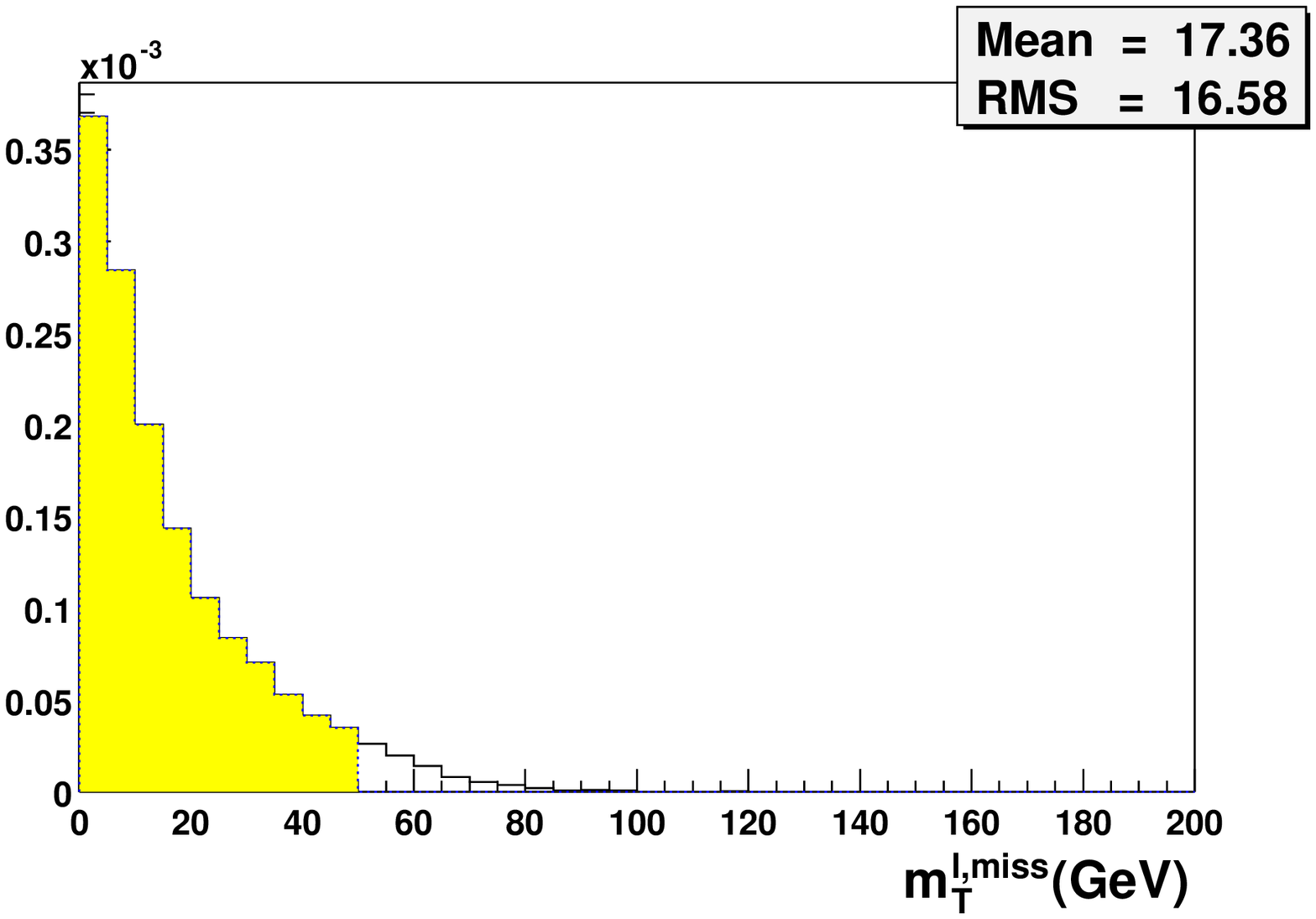,width=4.5cm, height=4.5cm}
     \epsfig{file=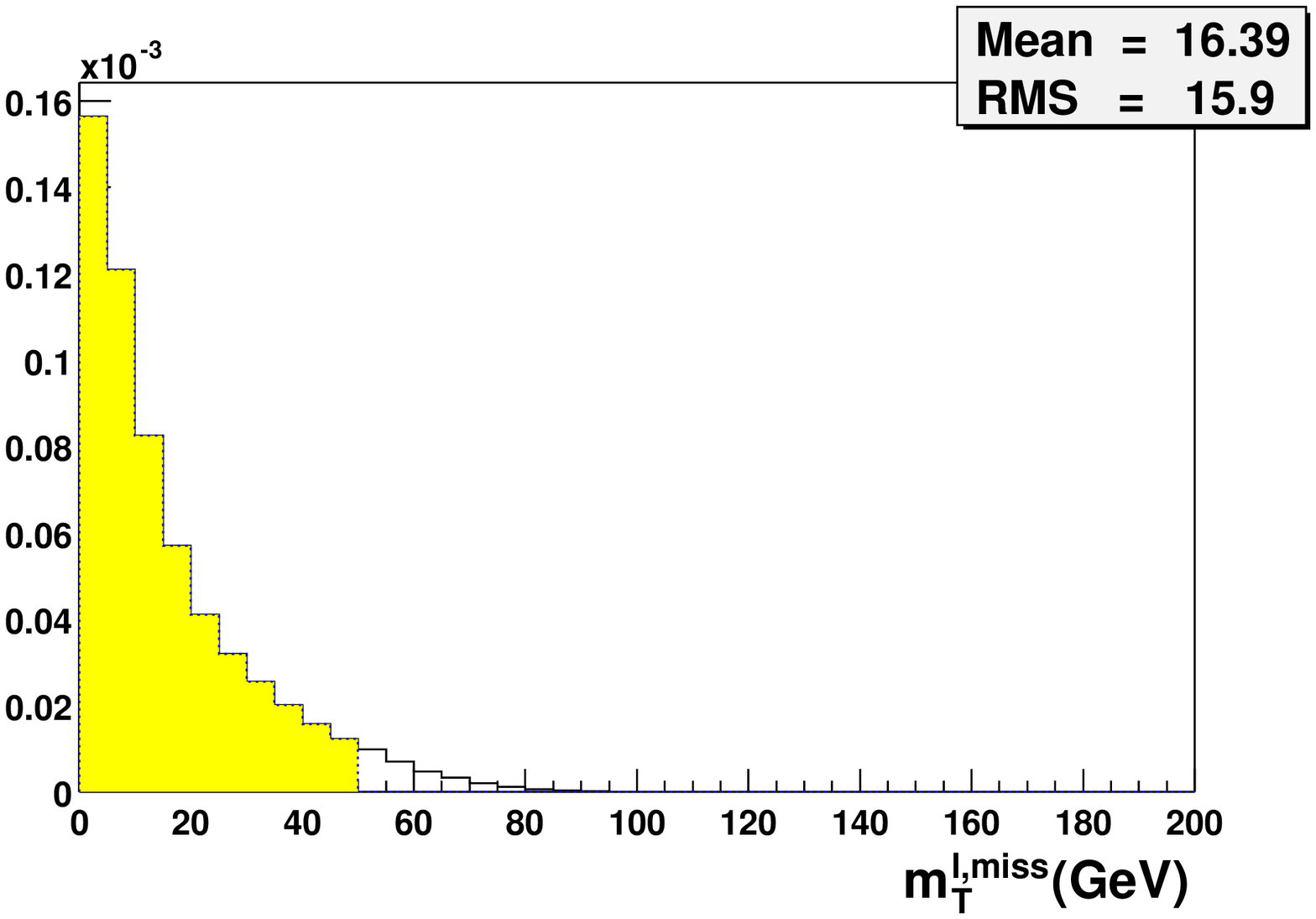,width=4.5cm, height=4.5cm}\\
}
{
     \epsfig{file=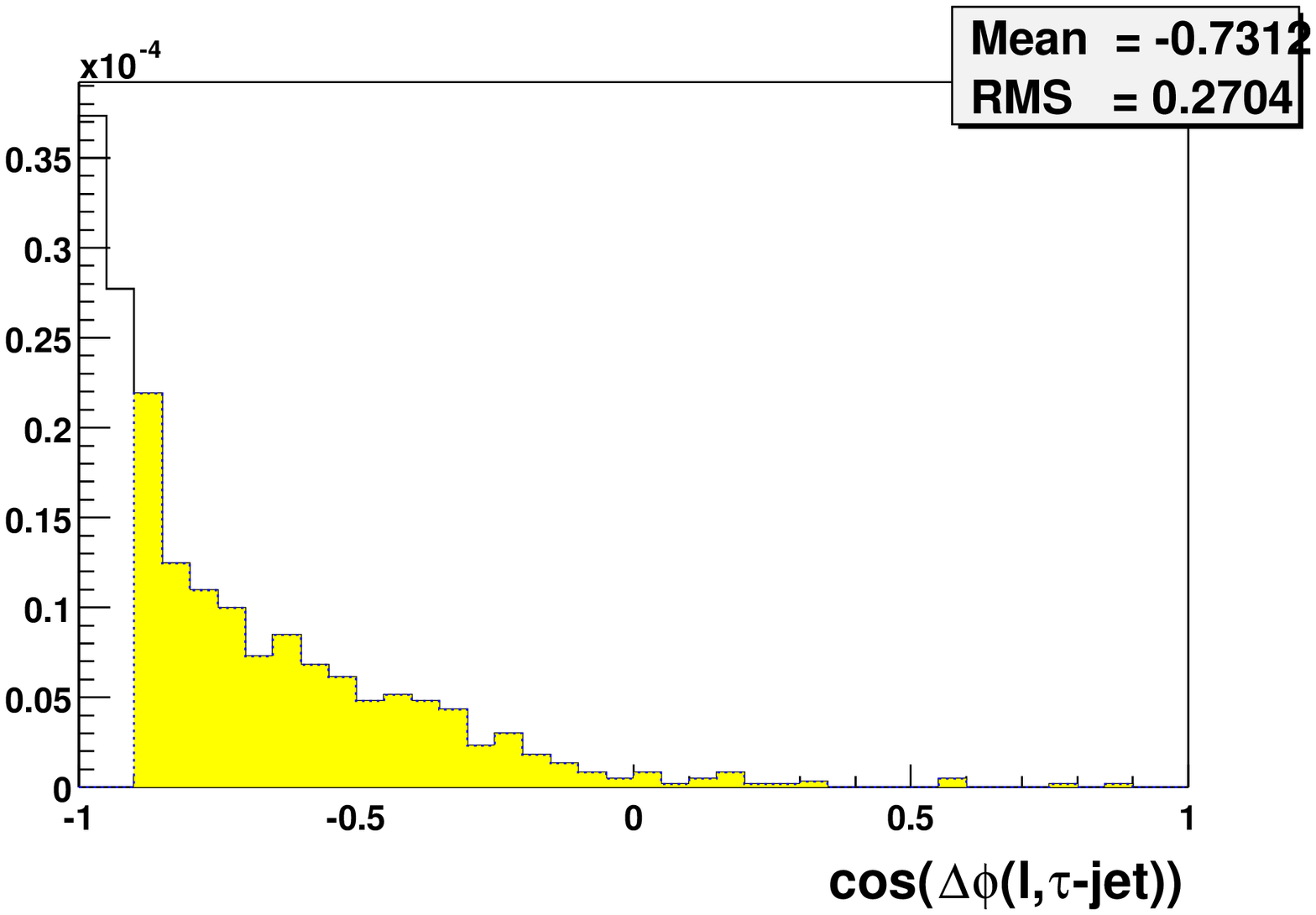,width=4.5cm, height=4.5cm}
     \epsfig{file=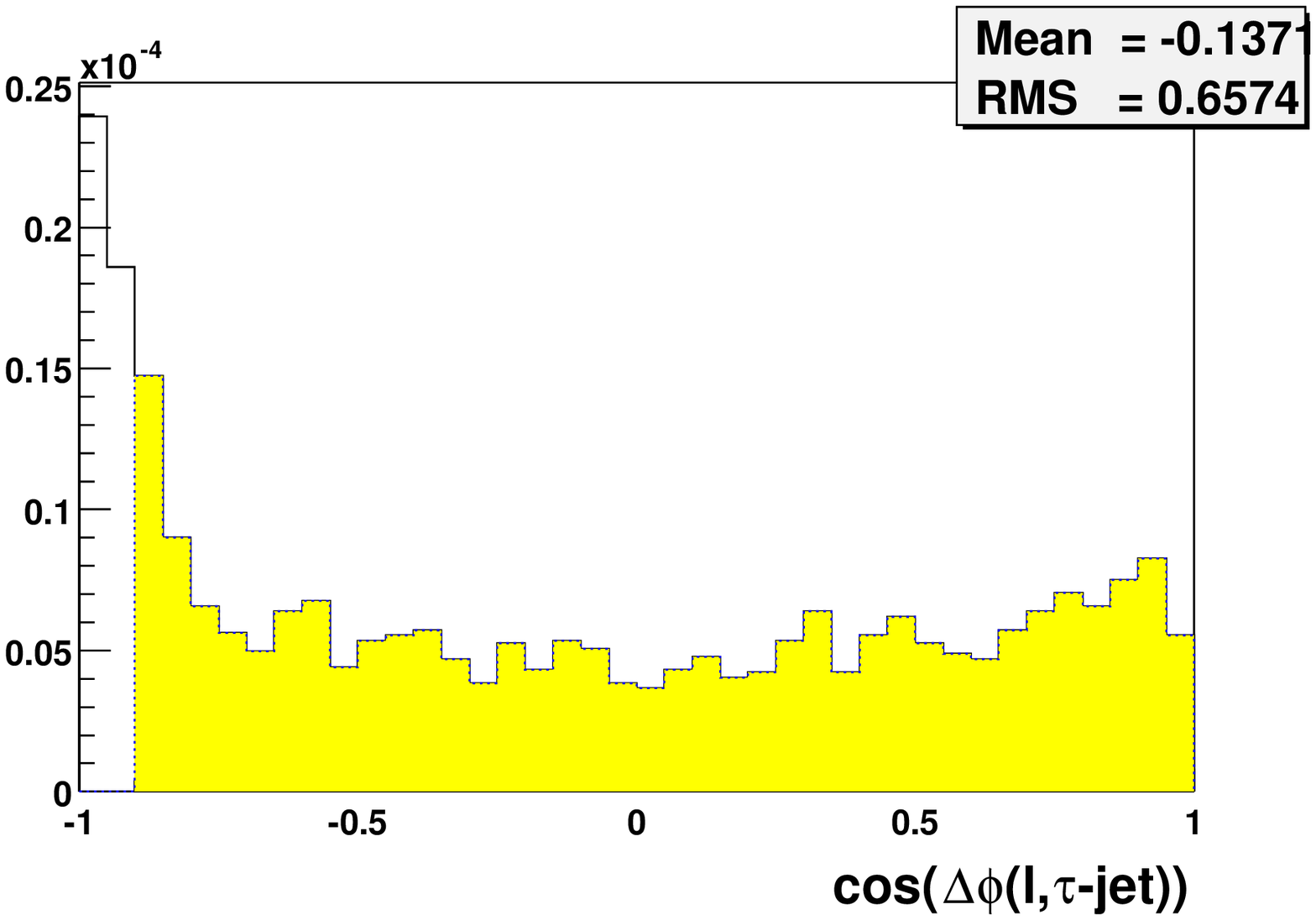,width=4.5cm, height=4.5cm}
     \epsfig{file=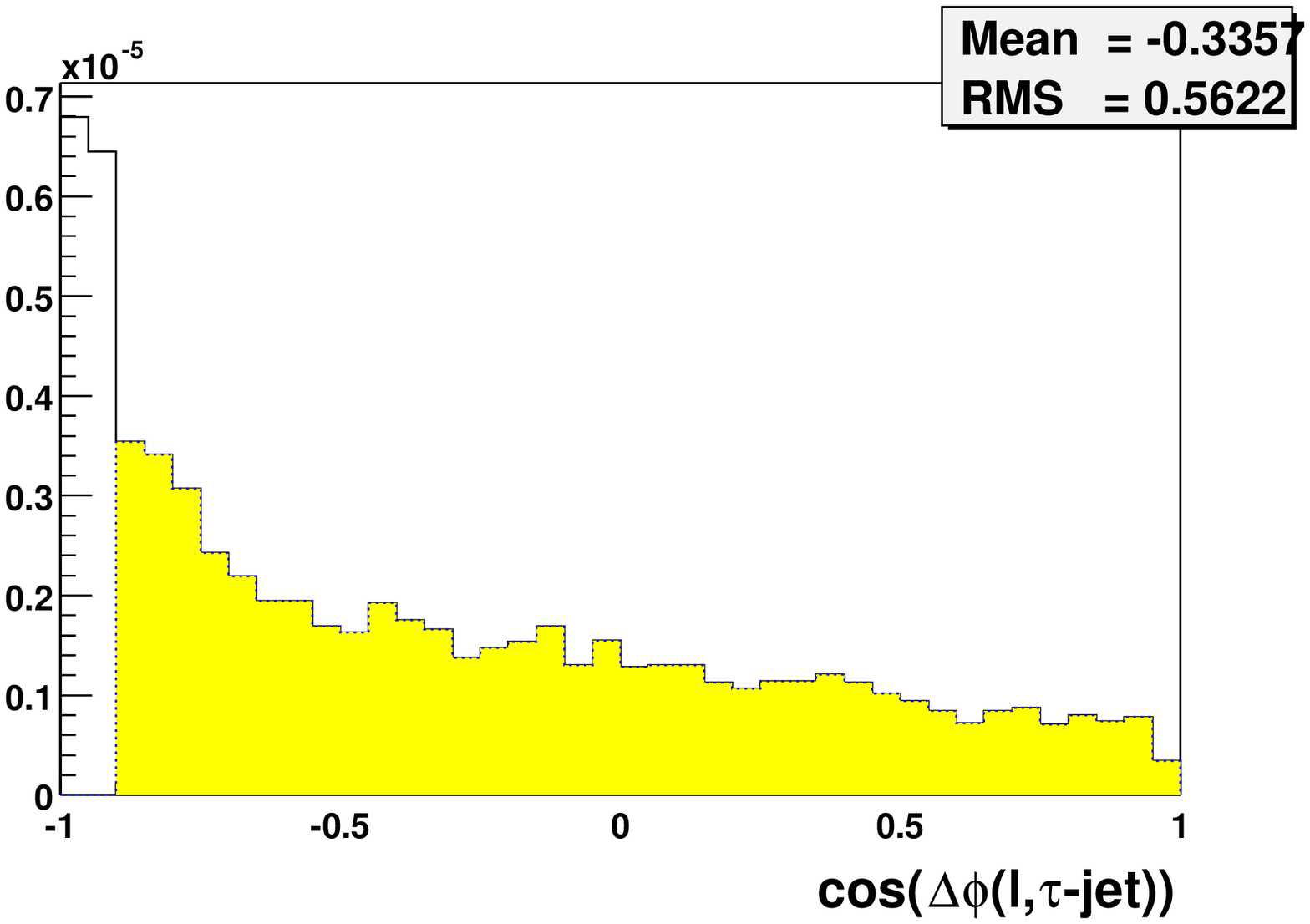,width=4.5cm, height=4.5cm}\\
}
\end{center}
\caption{\em
The characteristic kinematical distributions before respective selection, for 
different hard process used: $b \bar b \to H$ (left column), $gb
\to b  H$ (middle column) and $q \bar q \to b \bar b H$ (right
column). The shaded (yellow) area will be accepted by the selection.
Distributions normalised to total $\sigma \times BR$ [pb]. Every line of the plots
correspond to variables as used in cuts: lines 2, 3 and 6 of table \ref{TS6.1}.
Symmetric shape of the distributions of the first line of the plots provides
technical test of the simulation.   
\label{FS7.1}} 
\end{Fighere}
\vspace{2.5mm}

 Let us stress that results presented in Sections 
5 and 6 should be considered as illustration of the problem, and not as 
an {\it optimised expected performance} of the LHC experiments.

\begin{Tabhere} 
\newcommand{\lstrut}{{$\strut\atop\strut$}} 
\begin{center}
\begin{tabular}{|c||c|c|c||c|c|} \hline \hline
Selection & $ b \bar b \to H$ &  $ g b \to b H$ &   $ b \bar b H$ & $ gg \to H$ & $qqH$  \\
\hline \hline 
Basic selection &  & &  & &   \\
\hline \hline 
 1 iso $\ell$, $p_T^{\ell}>20$ GeV &  & &  & &  \\
  1 $\tau$-jet, $p_T^{\tau-jet}>30$ GeV & $ 19.5 \cdot 10^{-2}$ & $ 19.3 \cdot 10^{-2}$ & $  19.7 \cdot 10^{-2}$ & $ 19.5 \cdot 10^{-2}$  & $ 22.2 \cdot 10^{-2}$ \\
\hline \hline
$|sin(\Delta \phi_{\ell \ \tau - jet})| > 0.2$ & $ 9.6 \cdot 10^{-2}$  & $10.5 \cdot 10^{-2}$  & $  10.4 \cdot 10^{-2}$  & $ 10.9 \cdot 10^{-2}$ & $ 19.8 \cdot 10^{-2}$ \\
\hline
$m_T^{\ell, miss}< 50$ GeV & $ 8.9 \cdot 10^{-2}$   & $ 9.8 \cdot 10^{-2}$ &  $ 9.7 \cdot 10^{-2}$ & $ 10.1 \cdot 10^{-2}$ & $18.4 \cdot 10^{-2}$  \\
\hline
resolved neutrinos &  $ 5.5 \cdot 10^{-2}$  & $ 6.6 \cdot 10^{-2}$ & $  6.0 \cdot 10^{-2}$ & $ 7.6 \cdot 10^{-2}$& $15.3 \cdot 10^{-2}$ \\
\hline \hline 
Additional selection &  & &   \\ 
\hline \hline 
$p_T^{miss}> 30$ GeV &  $ 9.1 \cdot 10^{-3}$  &  $ 2.1 \cdot 10^{-2}$ & $   1.4 \cdot 10^{-2}$  & $ 3.0 \cdot 10^{-2}$ &$  10.1\cdot 10^{-2}$   \\
\hline
$cos(\Delta \phi_{\ell \ \tau -jet}) > -0.9$ &  $ 6.5 \cdot 10^{-3}$  &  $ 1.8\cdot 10^{-2}$ & $  1.2 \cdot 10^{-2}$ & $  2.7\cdot 10^{-2}$  & $ 9.8 \cdot 10^{-2}$ \\  
\hline
$R_{\ell \ \tau - jet} < 2.8$ &  $ 6.1  \cdot 10^{-3}$  & $ 1.7\cdot 10^{-2}$  & $  1.1  \cdot 10^{-2}$  & $ 2.6 \cdot 10^{-2}$ & $ 9.7 \cdot 10^{-2}$  \\  
\hline \hline 
\end{tabular}
\end{center}
  \caption {\em The cumulative acceptances of the selection criteria for different
approaches of modeling production process. For each subsequent line effect of the
additional cut off is added. 
\label{TS6.1}}  
\end{Tabhere}
\vspace{2.5mm}

\begin{Tabhere} 
\newcommand{\lstrut}{{$\strut\atop\strut$}}
\begin{center}
\begin{tabular}{|c||c|c|c||c|c|} \hline \hline
Selection & $ b \bar b \to H$ &  $g b \to b H$ & $ b \bar b H$ & $ gg \to H$ & $qqH$ \\
\hline \hline 
Basic selection & 18.3 GeV & 14.4 GeV &  16.8 GeV   & 12.8 GeV  & 10.0 GeV  \\
                & (62.2 \%)  &(65.5 \%) &  (66.0\%) & (71.5 \%) &  (82.4 \%) \\
\hline
$p_T^{miss}> 30$ GeV & 13.6 GeV & 11.4 GeV &  12.5 GeV   &  9.8 GeV & 9.0 GeV \\
                &  ( 62.5\%)  &( 72.6 \%) & ( 71.1\%) & ( 78.2 \%)&  (85.2 \%)  \\
\hline
$cos(\Delta \phi_{\ell \ \tau - jet}) > -0.9$ & 13.1 GeV & 11.1 GeV  & 11.9 GeV  &9.7  GeV   & 8.9 GeV   \\  
                &  ( 71.5\%)  &( 79.0 \%)  & ( 77.4\%) & ( 82.7 \%)  & (87.1\%) \\
\hline
$R_{\ell \ \tau - jet} < 2.8$ & 13.0 GeV & 11.1 GeV &  11.8  GeV & 9.7  GeV & 8.9 GeV   \\  
                &  ( 73.4\%)  &( 80.1 \%) &( 78.0\%) & (83.4 \%)  & ( 87.6 \%) \\
\hline \hline
\end{tabular}
\end{center}
  \caption {\em Gaussian resolution for the reconstructed $\tau \tau$ 
system for different approaches of modeling production process.
In brackets acceptance within mass window of $m_H~\pm$~20~GeV is shown.
\label{TS6.2}} 
\end{Tabhere}
\vspace{2.5mm}

\begin{Tabhere} 
\newcommand{\lstrut}{{$\strut\atop\strut$}}
\begin{center}
\begin{tabular}{|c||c|c|c||c|c|} \hline \hline
Selection & $ b \bar b \to H$ &  $g b \to b H$ & $ b \bar b H$ & $ gg \to H$ & $qqH$ \\
\hline \hline 
Basic selection &  $ 3.4 \cdot 10^{-2}$ &  $ 4.3 \cdot 10^{-2}$ &  $ 4.0 \cdot 10^{-2}$ &  $ 5.4 \cdot 10^{-2}$  &  $ 12.6 \cdot 10^{-2}$  \\
\hline
$p_T^{miss}> 30$ GeV &  $ 5.7 \cdot 10^{-3}$ &  $ 1.5  \cdot 10^{-2}$  &  $ 1.0 \cdot 10^{-2}$&  $ 2.3 \cdot 10^{-2}$&  $ 8.6 \cdot 10^{-2}$     \\
\hline
$cos(\Delta \phi_{\ell \ \tau-jet}) > -0.9$ &  $ 4.7 \cdot 10^{-3}$  &  $ 1.4 \cdot 10^{-2}$  &  $ 9.0  \cdot 10^{-3}$  &  $ 2.2  \cdot 10^{-2}$  &  $ 8.5  \cdot 10^{-2}$    \\  
\hline
$R_{\ell \ \tau-jet} < 2.8$ &  $ 4.5 \cdot 10^{-3}$ &   $ 1.4 \cdot 10^{-2}$  &   $ 8.8 \cdot 10^{-3}$&   $ 2.1 \cdot 10^{-2}$&   $ 8.5 \cdot 10^{-2}$ \\  
\hline \hline
\end{tabular}
\end{center}
  \caption {\em The cumulative acceptance in the mass window $m_H~\pm$~20~GeV,
for different approaches of modeling production process.
\label{TS6.3}} 
\end{Tabhere}
\vspace{2.5mm}

\section{Conclusions}

In the paper we have discussed one of the key  signatures at LHC:  $(\ell$~$\ell$~$p_T^{miss})$  
and $(\ell$~$\tau-$~jet~$p_T^{miss})$. They are important for searches of the
Higgs boson if it decays into $\tau$-lepton pair.
 Numerical results were  collected in sections 5 and 6.

We have concentrated on the discussion of the reconstruction efficiency and mass resolution
of the LHC experiments. We have varied the production processes, also, we have used
different methods for the simulation of the same channels, effectively re-shuffling parts 
of the hard
process into parton shower. In this way 
we have obtained different event topologies. Even though we have applied  rather simplified
analysis, for example we have neglected
additional cut-offs related to the background suppression and we have used simplified 
simulation for 
the detector response only, our analysis give an insight into the complexity of the problem. 
We have shown that 
the choice of production hard process affects the cumulative acceptance in the mass window for 
the signal reconstruction by a factor of few (even up to a factor of 10 for VBF fusion channel).
The main effect comes from interplay of average transverse momentum 
of the Higgs boson, simulated differently depending on the way how the hard process is chosen
and ambiguities in measurement of $p_T^{miss}$. 
This have an impact on the efficiency for finding physical solution of the neutrino system, 
necessary for the $m_{\tau \tau}$ reconstruction. Transverse momentum  determines  quality  of 
the resolution, in particular in the regions of tails.

Strong sensitivity on the production topology indicates that for the more complicated production 
processes, such as $b \bar b \to H$ Yukawa induced mechanism,  the possible large theoretical 
systematic error may still need to be assumed for the predictions, even if small theoretical 
uncertainties
obtained for  the overall normalisation can be nowdays well controlled
with NNLO calculations. For inclusive quantities effects related to the QCD regularisation 
and renormalisation scale are
well understood. However, the  fully exclusive Monte Carlo implementation is  mandatory for the experimentally 
useful analysis. To convince the reader, we have explicitely studied effects of principal selection cut-offs
which experiments will be most probably forced to use in data analysis.

Let us stress, that  we have not even touched the subject of the impact of those 
selection criteria which must  be applied to optimise rejection against expected
background (b-jet identification or b-jet veto). It will only add complexity to the discussion.
Asking for b-jet identification or b-jet veto will enhance contribution from some NLO or NNLO
terms, with respect to the total inclusive cross-section, in a manner which is difficult to 
predict
analytically, as predictions need to be convoluted with detector effects (jets reconstruction and jets 
identification).
Thus even stronger argument for the availabity 
of the existing theoretical calculations in  the form of exclusive Monte Carlo, suitable for more
educated studies than the one presented here. 
Without such a tool it will be difficult to decide in unambiguous way,  how much of the mentioned above, 
factor of 3 to 10, ambiguity on cumulative acceptance, 
indeed contributes to theoretical uncertainty for the
{\it realistic} signature of the Higgs boson.  Note, that ambiguity depends
drastically on how complete is the choice of selection cuts. If we had taken 
only first step (first lines of tables \ref{TS5.1} and \ref{TS6.1}) of the 
basic selection cuts, then we could conclude that acceptance is 
independent, both from different approaches to modelling of production process 
and from production process as well. Thus that it can be represented 
by an universal factor,  up to about $\pm 10 \%$ precision level. 
The conclusion would be, as we could see, overhelmingly false. On the other 
hand, turning the question around, one can ask, if experimental selections can 
be modified to diminish dependence on the topological feautures of the events 
(namely quality of reconstruction and absolute value of the $p_T^{miss}$). 


Let us note also, that for the Higgs boson at 120 GeV dominant background will come from the
irreducible Drell-Yan $Z/\gamma^* \to \tau \tau$ production. The signal will appear above 
steeply failing edge of the resonant peak at the nominal Z-boson mass, and will be dominated
by the misreconstructed on-shell Z-boson events. Calibration of  the background from the shape
outside signal mass window will be extremely difficult.

Finally,
let us  recall  again, that for the MSSM Higss scenarios in the parameter space corresponding
to the A/H/h masses in range of 100-200 GeV, one will have to combine distinct
production modes ($gg \to H$  and $b \bar b \to H$), and different final states of $h \to \tau^+ \tau^-$
with subsequent decays either into  lepton-lepton or lepton-hadron channels.  
The signatures of the Higgs bosons  h/H/A may overlap
(even if there is no mass degeneracy between h/H/A).
The goal of the experiment will be not only to establish evidence for the signal,
but also to measure properties of the model (eg. couplings) from the observed versus predicted
signal rates.
The key challenge for evaluating remaining theoretical systematic error 
will require not only the proper normalisation of total cross section 
but also modeling of the production mechanism including differential
distributions, essential as demonstrated here, for experimental signal reconstruction.

\section*{Acknowledgments}
One of the authors (ZW) is thankful to KITP Santa Barbara CA for inspiring 
atmosphere, which in particular, led  to  discussions with
F. Maltoni,  D. Rainwater  and T. Sj\"ostrand at a time when final steps 
of this work were completed. 
ERW is thankful to R.~Harlander,
F.~Maltoni, T.~Plehn, M.~Spira, S.~Willenbrock and D.~Zeppenfeld for several
discussions on the relative merits of the VFS and  FFS approaches
for the experimental analysis. 
ERW and TS express their gratitude for the stimulating atmosphere of the 
ATLAS Higgs Working group where this work was initiated.
\providecommand{\href}[2]{#2}
\end {document}